%% file: main.tex
\documentclass[reprint,amsmath,amssymb,aps,prb,superscriptaddress,nofootinbib,longbibliography]{revtex4-2}

\usepackage{graphicx}
\usepackage{natmove}
\usepackage{hyperref}
\hypersetup{colorlinks,allcolors=blue}

\usepackage{macros/my_formatting}
\usepackage{macros/my_macros}
\usepackage{macros/popupcite}

\input{main.bbl}
\begin{document}

\input{cap.tex}
\input{introduction.tex}
\input{methods.tex}
\input{results.tex}
\input{acknowledgements.tex}

\input{main.bbl}

\end{document}

%% file: cap.tex
\title{Ultimate sharpness of the tunneling resonance in vertical heterostructures}

\author{Georgy Alymov}
\email{alymov@phystech.edu}
\affiliation{Moscow Institute of Physics and Technology (National Research University), Dolgoprudny 141700, Russia}

\author{Dmitry Svintsov}
\affiliation{Moscow Institute of Physics and Technology (National Research University), Dolgoprudny 141700, Russia}
             
\input{abstract.tex}

\maketitle

%% file: abstract.tex
\begin{abstract}
  Heterostructures comprised of two two-dimensional electron systems (2DES) separated by a dielectric exhibit resonant tunneling when the band structures of both systems are aligned. It is commonly assumed that the height and width of the resonant peak in the tunneling current is determined
  by electron scattering and rotational misalignment of crystal structures of the 2DES. We identify two fundamental factors limiting the maximum height and steepness of the resonance: coupling to
  contacts and tunnel splitting of energy levels. The upper limit of the tunneling
  current is the number of electrons available for tunneling times half the tunnel coupling between the 2DES. As a result of a tradeoff between the contact-induced level broadening and contact resistance, the maximum current is only achievable when the coupling to contacts equals the tunnel
  level splitting. According to our model calculations, the limiting behavior can be observed in double-gated graphene/few-layer hexagonal boron nitride/graphene heterostructures.
\end{abstract}

%% file: introduction.tex
\section{Introduction}
Tunneling heterostructures based on parallel two-dimensional electron systems (2DES) have numerous applications in electronics and photonics. The most established are the quantum well photodetectors~\cite{QWIPs} and quantum cascade lasers~\cite{QCL}, while rich prospects are anticipated for steep-switching tunneling transistors~\cite{Wang-vdW_TFET,Xiong-vdW_TFET}. From a fundamental viewpoint, experimental studies of tunneling current carry the information about electron-electron scattering~\cite{lifetime_by_tun_spectrocopy_Murphy,tunnel_spectroscopy_Turner,Spectral_function_resolved_by_tunneling,BLG-WSe2_quantum_lifetime} and correlations~\cite{Strong_correlations_tunneling,Tunneling_in_quantum_Hall_bilayers}. Recent advent of van der Waals heterostructures has considerably enriched the field of tunneling electronics~\cite{Gr-hBN_tunneling_model_2,Vdovin_e-ph,Greenaway_Tunnel_spectroscopy,Wallbank_tuning_valley_with_tunneling}.

The phenomenon of tunneling resonance between parallel 2DES is at the heart of these applications and fundamental studies, and understanding its ultimate magnitude and steepness plays a critical role. The resonance occurs upon perfect alignment of electron bands in the adjacent 2DES. Under this resonance, the charge carriers can transfer from one 2DES to another with conservation of in-plane momentum ${\bf k}$ and energy $\epsilon$. The simplest and most commonly used model predicting the resonant peak in current-voltage curves $J(V)$ is based on Bardeen's transfer Hamiltonian~{\cite{tunneling_Gr_coherence_length}}
\begin{eq}{golden_intro}
  J = \frac{2 \pi g e}{\hbar} \int  \frac{\mathd^2 \mathbf{k}}{(2 \pi)^2}
  (f_{b\mathbf{k}} - f_{t\mathbf{k}}) | v_{t b\mathbf{k}} |^2 \delta
  (\epsilon_{t\mathbf{k}} - \epsilon_{b\mathbf{k}}),
\end{eq}
where $f_{b/t\mathbf{k}}$ are the distribution functions in bottom (b) and top (t) layers, $\epsilon_{b/t\mathbf{k}}$ are the respective ${\bf k}$-dependent energy bands, and $v_{t b\mathbf{k}}$ is the tunneling matrix element.

When the emitter and collector bands are aligned, $\epsilon_{t\mathbf{k}} = \epsilon_{b\mathbf{k}}$, the transfer Hamiltonian approach in the form of Fermi's golden rule (\ref{eq:golden_intro}) predicts an infinite current~{\cite{Gr_tunneling_theory_review}}. This divergence is usually cured by incorporating electron scattering
{\cite{tunneling_disorder_sqrt_divergence,Gr-hBN_tunneling_model_3}}. 
Moreover, the assumed link between the resonance width and height and the inelastic quasiparticle scattering rate $\gamma_{\rm sc}$ became an experimental tool for studies of electron scattering~\cite{lifetime_by_tun_spectrocopy_Murphy,BLG-WSe2_quantum_lifetime}. 

The theory of tunneling between parallel 2DES in the form of scattering-broadened Fermi's golden rule cannot be considered satisfactory in the limit of weak scattering. Indeed, an electron placed into one of the two tunnel-coupled quantum wells has a finite 'tunneling frequency' $\tau^{-1}_{\rm tun} = v_{tb {\bf k}}/\hbar$. The coherent transfer between the wells cannot be faster than $\tau_{\rm tun}$, thus the resonant current density should be somehow limited by the product of the electron density per layer and the tunneling frequency. At the same time, the tunnel coupling alone cannot result in an interwell dc current, as the electron exhibits coherent beatings between the wells. Only the electron decoherence may interrupt the quantum beatings and lead to a finite current.

\begin{fig}{setup}{figs/setup}
  Schematic of vertical tunneling setup under consideration. Top and bottom two-dimensional systems (orange) are separated by a few-atom thick tunnel barrier (green). The 2DES are connected to the source (S) and drain (D) reservoirs with characteristic carrier exchange rate $\gamma$.
\end{fig}

In this paper, we analytically address the problem of maximum resonant tunneling current between parallel 2DES, accurately taking into account the tunnel coupling and decoherence induced by the presence of contacts. The experimental setup under consideration is shown in Fig.~\ref{fig:setup}. We analytically show that a combination of these factors determines the ultimate width and height of tunnel resonance even in the absence of disorder and carrier-carrier scattering. 
We show that even when scattering is negligible, the resonant peak in the tunneling $J(V)$
characteristics still has finite height and width limited by level splitting
in the tunnel-coupled emitter and collector and by coupling to the contacts with characteristic rate $\gamma$. The resonance height attains its maximum value when these couplings are matched, $\gamma=2|v_{tb}|$. We rigorously derive a generalization of golden-rule expression for the tunneling current accounting for these effects, and apply it to the experimentally relevant heterostructures comprised of graphene layers separated by hexagonal boron nitride (hBN). The effects of finite tunnel level splitting on the resonance height and width become pronounced for few-layer hBN barriers even if the graphene layers are rotationally misaligned.

Our derivation is based on the non-equilibrium Green's function formalism in the tight-binding approximation for interlayer coupling~{\cite{Datta}}. Each 2DES is coupled at its edge to a large electronic reservoir with carrier exchange rate $\gamma$. Though similar models have been applied to the transport in van der Waals heterostructures~{\cite{Gr-hBN_tunneling_model, Gr-hBN_tunneling_model_2, Gr-hBN_tunneling_model_3, Gr-hBN_tunneling_model_e-e, Gr-hBN_tunneling_model_4}}, the limitations on the resonance width and height imposed by tunnel splitting and coupling to the contacts remained undisclosed. 
The interplay between coupling to the contacts $\gamma$ and tunnel coupling is well-studied for quantum dot structures~\cite{multidot_Gurvitz,multidot_Nazarov,double_dot_Schoell,double_dot_finite_T,nonmonotonic_J_vs_gamma_3}. A similar interplay takes place in superlattices~{\cite{superlattice_Kazarinov,superlattice_Capasso}} and multi-well resonant tunneling diodes with bulk emitters and collectors~{\cite{multiwell_Gurwitz,double_well_Wingreen,double_well_Nagase,double_well_exp_splitting}}.
The non-monotonic dependence of current on the lead-channel coupling $\gamma$ was also reported in tight-binding simulations of in-plane transport {\cite{nonmonotonic_J_vs_gamma_1,nonmonotonic_J_vs_gamma_2}}.

%% file: methods.tex
\section{Methods}
\subsection{Model}

We consider a heterostructure formed of two 2D materials separated by a
dielectric. Hereafter, we will refer to them as top ($t$) and bottom ($b$)
leads.

Each lead is coupled to an infinite electron reservoir. In the reservoirs,
electrons obey Fermi-Dirac distributions $f_t (E)$, $f_b (E)$ with different
chemical potentials $\mu_t$, $\mu_b$. The bottom reservoir is grounded, $\mu_b
= 0$, and a bias voltage $V_{\tmop{bias}}$ is applied to the top reservoir,
$\mu_t = - e V_{\tmop{bias}}$.

The (single-particle) Hamiltonian of the whole system is
\begin{equation}
  \hat{H} = \left(\begin{array}{ccccc}
    \hat{H}_{t r} & \hat{T}_{t r}^{\dag} & 0 & 0 & 0\\
    \hat{T}_{t r} & \hat{H}_t & \hat{T}_t^{\dag} & 0 & 0\\
    0 & \hat{T}_t & \hat{H}_d & \hat{T}_b  & 0\\
    0 & 0 & \hat{T}_b^{\dag} & \hat{H}_b & \hat{T}_{b r} \\
    0 & 0 & 0 & \hat{T}_{b r}^{\dag} & \hat{H}_{b r}
  \end{array}\right), \label{eq:Hamiltonian}
\end{equation}
where $\hat{H}_{t (b)}$ is the Hamitonian of the top (bottom) 2D material,
$\hat{H}_{t (b) r}$ is the Hamitonian of the top (bottom) electron reservoir,
$\hat{H}_d$ is the Hamiltonian of the dielectric, and the rest of $\hat{H}$
describes electron hopping between different materials.

\subsection{Green's functions and eigenstates}

The retarded Green's function of the system is given by
\begin{equation}
  \hat{G}^R = \frac{1}{E - \hat{H} + i 0}, \label{eq:GR}
\end{equation}
and its leads block is
\begin{eq}{GR_leads}
  \hat{G}^R_l &= \left(\begin{array}{cc}
    \hat{G}^R_{t t} & \hat{G}^R_{t b}\\
    \hat{G}^R_{b t} & \hat{G}^R_{b b}
  \end{array}\right)\\ &= \left(\begin{array}{cc}
    E - \hat{H}_t - \hat{v}_{t t} - \hat{\sigma}_t & - \hat{v}_{t b}\\
    - \hat{v}_{b t} & E - \hat{H}_b - \hat{v}_{b b} - \hat{\sigma}_b
  \end{array}\right)^{- 1},
\end{eq}
where $\hat{v}_{i j} = \hat{T}^{\dag}_i \hat{g}^R_d \hat{T}_j$ and
$\hat{\sigma}_i = \hat{T}_{i r} \hat{g}^R_{i r} \hat{T}_{i r}^{\dag}$ are the
self-energies due to the lead-dielectric and lead-reservoir coupling,
respectively, and $\hat{g}^R$ denotes the retarded Green's functions of
isolated subsystems ($\hat{g}^R_d = \frac{1}{E - \hat{H}_d + i 0}$, etc.).

The diagonal terms $\hat{v}_{t t}$, $\hat{v}_{b b}$ describe the
dielectric-induced renormalization of the leads' bandstructures and will be
neglected. The tunneling matrix elements $\hat{v}_{t b}$, $\hat{v}_{b t}$
describe level splitting due to the tunnel coupling between leads. $\hat{v}_{b
t} = \hat{v}_{t b}^{\dagger}$ within the bandgap of the dielectric.

The main role of the reservoirs is to allow electrons to go in and out of
the heterostructure, resulting in a finite spectral broadening. We model it with a
single scalar quantity $\gamma$ ($\hat{\sigma}_t = \hat{\sigma}_b = - \frac{i
\gamma}{2}$).

In the single band model with electron dispersion relation
$\epsilon_{t (b) \mathbf{k}}$ inside the top (bottom) lead and assuming
momentum conservation,
\begin{equation}
  \hat{G}^R_{l\mathbf{k}} = \frac{1}{(E - \tilde{\epsilon}_{t\mathbf{k}}) (E -
  \tilde{\epsilon}_{b\mathbf{k}}) - | v_{t b\mathbf{k}} |^2}
  \left(\begin{array}{cc}
    E - \tilde{\epsilon}_{b\mathbf{k}} & v_{t b\mathbf{k}}\\
    v_{b t\mathbf{k}} & E - \tilde{\epsilon}_{t\mathbf{k}}
  \end{array}\right), \label{eq:GR_single-band}
\end{equation}
where $\mathbf{k}$ is the electron momentum, and $\tilde{\epsilon}_{t (b)
\mathbf{k}} = \epsilon_{t (b) \mathbf{k}} - \frac{i \gamma}{2}$ are the
(complex) electron energies renormalized by interaction with the reservoirs.

$\hat{G}^R_{l\mathbf{k}} (E)$ has poles at energies
\begin{eq}{split_levels}
  \tilde{\epsilon}_{\pm, \mathbf{k}} & = \epsilon_{\pm, \mathbf{k}} -
  \frac{i \gamma}{2}, \\
  \epsilon_{\pm, \mathbf{k}} & = \frac{\epsilon_{t\mathbf{k}} +
  \epsilon_{b\mathbf{k}}}{2} \pm \sqrt{\left( \frac{\epsilon_{t\mathbf{k}} -
  \epsilon_{b\mathbf{k}}}{2} \right)^2 + | v_{t b\mathbf{k}} |^2}, 
\end{eq}
corresponding to the eigenstates of the tunnel-coupled leads.

\subsection{Tunneling current}

The tunneling current can be evaluated using the Landauer--Caroli formula
{\cite{Datta,Caroli,Gr-hBN_tunneling_model_3,Gr-hBN_tunneling_model_4}}
\begin{equation}
  J = \frac{g e}{h} \int  \frac{\mathd^2 \mathbf{k}}{(2 \pi)^2} \int \mathd E
  (f_b - f_t) {\gamma G^R_{t b\mathbf{k}}}  {\gamma G^A_{b t\mathbf{k}}} ,
  \label{eq:Landauer-Caroli}
\end{equation}
where $g$ takes into account spin and valley degeneracy.

Putting Green's function (\ref{eq:GR_single-band}) into
(\ref{eq:Landauer-Caroli}), we arrive at
\begin{eq}{golden_rule_spectral_functions}
  J & = \frac{g e}{h} \int  \frac{\mathd^2 \mathbf{k}}{(2 \pi)^2} \int
  \mathd E (f_b - f_t) \gamma^2\\ &\times \left| \frac{v_{t b\mathbf{k}}}{\left( E -
  \epsilon_{+, \mathbf{k}} + \frac{i \gamma}{2} \right) \left( E -
  \epsilon_{-, \mathbf{k}} + \frac{i \gamma}{2} \right)} \right|^2 \\
  & = \frac{2 \pi g e}{\hbar} \int  \frac{\mathd^2 \mathbf{k}}{(2 \pi)^2}
  \int \mathd E (f_b - f_t) | v_{t b\mathbf{k}} |^2\\ &\times \Lambda_{\gamma} (E -
  \epsilon_{+, \mathbf{k}}) \Lambda_{\gamma} (E - \epsilon_{-, \mathbf{k}}),
\end{eq}
where $\Lambda_{\gamma} (E) \equiv \frac{1}{\pi} \frac{\gamma / 2}{E^2 +
\gamma^2 / 4}$ is a Lorentzian of width $\gamma$.

The product $\Lambda_{\gamma} (E - \epsilon_{+, \mathbf{k}}) \Lambda_{\gamma}
(E - \epsilon_{-, \mathbf{k}})$ has two peaks with equal weights located near
$\epsilon_{+, \mathbf{k}}$, $\epsilon_{-, \mathbf{k}}$. If the density of
states and the tunneling matrix element do not change significantly on the
energy scale of $\gamma$ (note that the dependence $v_{t b\mathbf{k}} (E)$ in
layered materials is not exponential, but power-law
{\cite{power-law_T_vs_E}}), we can approximate $[f_b (E) - f_t (E)] | v_{t
b\mathbf{k}} (E) |^2$ by the average of its values at $\epsilon_{\pm,
\mathbf{k}}$ and integrate over energy analytically:
\begin{equation}
  J = \frac{2 \pi g e}{\hbar} \int  \frac{\mathd^2 \mathbf{k}}{(2 \pi)^2}
  (f_{b\mathbf{k}} - f_{t\mathbf{k}}) | v_{t b\mathbf{k}} |^2 \Lambda_{2
  \gamma} (\epsilon_{+, \mathbf{k}} - \epsilon_{-, \mathbf{k}}),
  \label{eq:broadened_golden_rule}
\end{equation}
where $(f_{b\mathbf{k}} - f_{t\mathbf{k}}) | v_{t b\mathbf{k}} |^2$ is a
shorthand for $\frac{1}{2} \sum_{\pm} [f_b (\epsilon_{\pm, \mathbf{k}}) - f_t
(\epsilon_{\pm, \mathbf{k}})] | v_{t b\mathbf{k}} (\epsilon_{\pm, \mathbf{k}})
|^2$.

When the bandstructures of the top and bottom lead are significantly
mismatched, such as in the cases of rotationally misaligned
structures, interband tunneling, or leads made of different materials,
spectral broadening and energy shifts due to tunnel coupling are usually
unimportant, and eq. (\ref{eq:broadened_golden_rule}) reduces to the usual
Fermi's golden rule
\begin{equation}
  J = \frac{2 \pi g e}{\hbar} \int  \frac{\mathd^2 \mathbf{k}}{(2 \pi)^2}
  (f_{b\mathbf{k}} - f_{t\mathbf{k}}) | v_{t b\mathbf{k}} |^2 \delta
  (\epsilon_{t\mathbf{k}} - \epsilon_{b\mathbf{k}}) . \label{eq:golden_rule}
\end{equation}
The conditions of its validity are: (i) tunneling is allowed by conservation
laws (nonzero current by Fermi's golden rule), (ii) negligible contact
resistance ($\gamma \gg | v_{t b\mathbf{k}} |$), (iii) negligible spectral
broadening ($\gamma$ is much smaller than the characteristic energy mismatch).

On the other hand, in the case of perfect alignment 
($\epsilon_{t\mathbf{k}} = \epsilon_{b\mathbf{k}}$) Fermi's golden
rule predicts infinite current
{\cite{Gr_tunneling_theory_review,tunneling_disorder_sqrt_divergence,Gr-hBN_tunneling_model_3,tunneling_Gr_correlated_disorder_divergence}},
while eq. (\ref{eq:broadened_golden_rule}) gives a finite answer:
\begin{eq}{verbose_resonance_height}
  J = && \frac{2 \pi g e}{\hbar} & \int  \frac{\mathd^2 \mathbf{k}}{(2 \pi)^2}
  (f_{b\mathbf{k}} - f_{t\mathbf{k}}) | v_{t b\mathbf{k}} |^2 \Lambda_{2
  \gamma} (2 | v_{t b\mathbf{k}} |) \\
   = && \frac{g e}{\hbar} & \int  \frac{\mathd^2 \mathbf{k}}{(2 \pi)^2}
  (f_{b\mathbf{k}} - f_{t\mathbf{k}}) \frac{\gamma}{2} \frac{| v_{t
  b\mathbf{k}} |^2}{| v_{t b\mathbf{k}} |^2 + \gamma^2 / 4} . 
\end{eq}

Tunneling between perfectly or nearly perfectly aligned bands in van der Waals
heterostructures is usually referred to as resonant tunneling \cite{Gr-hBN_tunneling_model_2}. 
This is a different, but closely related concept to the usual resonant tunneling in 
double-barrier heterostructures \cite{Datta}.

\begin{fig}{height-width}{figs/height-width}
  The height $J$ and width $w$ of the resonant peak in the $I$-$V$ characteristics of a van der Waals heterostructure vs the coupling to contacts $\gamma$ in the case of negligible scattering and rotational misalignment. $J_{\rm max}$ is the maximum possible height, $w_{\rm min}$ is the minimum possible width, $|v_{tb}|$ is the tunneling matrix element.
\end{fig}

Equation (\ref{eq:verbose_resonance_height}) can be written in a compact form:
\begin{equation}
  J = \frac{e}{\hbar} n_{\tmop{tun}} \frac{\gamma}{2} \frac{| v_{t b} |^2}{|
  v_{t b} |^2 + \gamma^2 / 4}, \label{eq:resonance_height}
\end{equation}
where $| v_{t b} |$ is a suitably averaged tunneling matrix element, and
\begin{equation}
  n_{\tmop{tun}} = g \int  \frac{\mathd^2 \mathbf{k}}{(2 \pi)^2}
  (f_{b\mathbf{k}} - f_{t\mathbf{k}}) \label{eq:tunneling_density}
\end{equation}
is the number of electrons participating in tunneling.

The peak current (\ref{eq:resonance_height}) depends on the coupling $\gamma$
between the leads and reservoirs and vanishes if $\gamma$ is either very small
or very large (blue curve in Fig.~\ref{fig:height-width})
{\cite{nonmonotonic_J_vs_gamma_1,nonmonotonic_J_vs_gamma_2,nonmonotonic_J_vs_gamma_3}}.
If $\gamma \ll | v_{t b} |$, the lead-reservoir coupling becomes the main
bottleneck in electron flow and the current is dominated by the contact
resistance. If $\gamma \gg | v_{t b} |$, the spectral weight of
electrons becomes spread over a large energy window, thereby reducing the
tunneling current.

The tunneling current at the resonance attains its maximum possible value at
$\gamma = 2 | v_{t b} |$:
\begin{equation}
  J_{\max} = e n_{\tmop{tun}} \frac{| v_{t b} |}{2 \hbar}
  \label{eq:ultimate_current} .
\end{equation}
In the context of tunneling through quantum dots, this non-monotonic $J
(\gamma)$ dependence has been described as a quantum transport analog of
Kramers' turnover in the rates of chemical reactions
{\cite{nonmonotonic_J_vs_gamma_1,lonely_chemical_reference}} or a
manifestation of the quantum Zeno effect {\cite{multidot_Nazarov}}.

\subsection{Relation to Fermi's golden rule}

\begin{fullfig}{stacking}{figs/stacking}
  (a) Relative orientation of the graphene and
  hBN layers with respect to the chosen coordinate system. Atoms in bottom
  layers are drawn bigger. (b) Brillouin zones of the top (thin lines) and
  bottom (thick lines) graphene. Circles illustrate the Fermi surfaces. Labels
  indicate $K$ and $K'$ valleys; equivalent Dirac cones are numbered with
  subscripts (index $n$ in eq. (\ref{eq:Dirac_points_momenta})). Rotational
  misalignment and lattice mismatch between graphene and hBN are exaggerated
  for clarity.
\end{fullfig}

We have obtained an expression for the height of the resonant peak (\ref{eq:resonance_height}),
 and now we proceed to calculate its width.

According to eqs. (\ref{eq:broadened_golden_rule}) and
(\ref{eq:split_levels}), the tunnel level splitting and coupling to the
reservoirs replace the delta function in Fermi's golden rule
(\ref{eq:golden_rule}) with a Lorentzian of width $2 \sqrt{\gamma^2 + 4 | v_{t
b \mathbf{k}} |^2}$:
\begin{eq}{attenuated_broadened_golden_rule}
  J = \frac{2 \pi g e}{\hbar} &\int  \frac{\mathd^2 \mathbf{k}}{(2 \pi)^2}
  \frac{\gamma}{\sqrt{\gamma^2 + 4 | v_{t b\mathbf{k}} |^2}}\\ &\times (f_{b\mathbf{k}}
  - f_{t\mathbf{k}}) | v_{t b\mathbf{k}} |^2 \Lambda_{2 \sqrt{\gamma^2 + 4 |
  v_{t b\mathbf{k}} |^2}} (\epsilon_{t\mathbf{k}} - \epsilon_{b\mathbf{k}}) .
\end{eq}
This spectral broadening translates into a corresponding broadening of the
resonant peak, as we will now show for the case of momentum- and
energy-independent $| v_{t b} |$.

Suppose the electric potentials of the leads $\varphi_t$, $\varphi_b$ can be
controlled by the voltage $V_{\tmop{bias}}$ applied between the leads and/or,
in gated structures, by the gate voltage. An external electric field will
shift the energies of all electron states in the top (bottom) lead by $U_{t
(b)} = - e \varphi_{t (b)}$.

To recover Fermi's golden rule, we insert an auxiliary delta function into
(\ref{eq:attenuated_broadened_golden_rule}):
\begin{eq}{convolved_golden_rule}
  J &= \frac{\gamma}{\sqrt{\gamma^2 + 4 | v_{t b} |^2}} \int d \delta E
  \Lambda_{2 \sqrt{\gamma^2 + 4 | v_{t b} |^2}} (\delta E)\\ &\times\frac{2 \pi g
  e}{\hbar} \int  \frac{\mathd^2 \mathbf{k}}{(2 \pi)^2} (f_{b\mathbf{k}} -
  f_{t\mathbf{k}}) | v_{t b} |^2 \delta (\epsilon_{t\mathbf{k}} -
  \epsilon_{b\mathbf{k}} - \delta E) .
\end{eq}

Now the integrand has almost the form of Fermi's golden rule, except there is
an energy mismatch $\delta E$ and the distribution functions are taken at
energies $\epsilon_{\pm, \mathbf{k}}$. This can be fixed with appropriate
shifts of the leads' bandstructures:
\begin{eq}{band_shifts}
  \epsilon_{t\mathbf{k}, \tmop{shifted}} & = \epsilon_{t\mathbf{k}} -
  \frac{\delta E}{2} \pm \sqrt{\frac{\delta E^2}{4} + | v_{t b} |^2} =
  \epsilon_{\pm \mathbf{k}}, \\
  \epsilon_{b\mathbf{k}, \tmop{shifted}} & = \epsilon_{b\mathbf{k}} +
  \frac{\delta E}{2} \pm \sqrt{\frac{\delta E^2}{4} + | v_{t b} |^2} =
  \epsilon_{\pm \mathbf{k}} .
\end{eq}

Thus, the tunneling current $J$ is related to the current $J_0$ computed via
Fermi's golden rule with different electrical potentials of the leads:
\begin{eq}{broadened_vs_sharp_current}
  J (U_{t,} U_b) & = \frac{\gamma}{\sqrt{\gamma^2 + 4 | v_{t b} |^2}} \int d
  \delta E \Lambda_{2 \sqrt{\gamma^2 + 4 | v_{t b} |^2}} (\delta E)\\
  & \times \frac{1}{2} \sum_{\pm} J_0 \left( U_t - \frac{\delta E}{2} \pm
  \sqrt{\frac{\delta E^2}{4} + | v_{t b} |^2}, \right.\\& \phantom{\times \frac{1}{2} \sum_{\pm} J_0 \left( \right.} \left. U_b + \frac{\delta E}{2} \pm
  \sqrt{\frac{\delta E^2}{4} + | v_{t b} |^2} \right) . 
\end{eq}

If the current depends on the band alignment much stronger than on the Fermi
level, eq. (\ref{eq:broadened_vs_sharp_current}) simplifies to
\begin{eq}{convolved_current}
  J (\Delta U_{t b}) &= \frac{\gamma}{\sqrt{\gamma^2 + 4 | v_{t b} |^2}}\\ &\times \int d
  \delta E \Lambda_{2 \sqrt{\gamma^2 + 4 | v_{t b} |^2}} (\delta E) J_0
  (\Delta U_{t b} - \delta E),
\end{eq}
where $\Delta U_{t b} = U_t - U_b$. At fixed Fermi energies $E_{F t}$, $E_{F
b}$ in the leads, $\Delta U_{t b} = - e V_{\tmop{bias}} - E_{F t} + E_{F b}$.

Therefore, the tunneling $I$-$V$ curves can be obtained from Fermi's golden
rule calculations by convolving the results with a Lorentzian of width $2
\sqrt{\gamma^2 + 4 | v_{t b} |^2} / e$ and multiplying them by an extra
attenuation factor $\frac{\gamma}{\sqrt{\gamma^2 + 4 | v_{t b} |^2}}$. This is
especially useful when the Fermi's golden rule resuts are known analytically.

As a consequence, the width of the resonant peak in the $I$-$V$ curves is $2
\sqrt{\gamma^2 + 4 | v_{t b} |^2} / e$ and cannot be smaller than $4| v_{t b} | / e$---twice the
tunnel level splitting (orange curve in Fig.~\ref{fig:height-width}).

Equations similar to (\ref{eq:resonance_height}),
(\ref{eq:attenuated_broadened_golden_rule}) have been obtained in the context
of resonant tunneling through coupled quantum dots
{\cite{multidot_Gurvitz,multidot_Nazarov,double_dot_Schoell,double_dot_finite_T}}
and superlattices {\cite{superlattice_Kazarinov,superlattice_Capasso}}. In the
latter case, carriers are provided by excited subbands instead of contacts.

\subsection{Graphene/\MakeLowercase{h}BN/graphene heterostructure}

To put our results into a practical perspective, we apply the developed theory
to a graphene/few-layer hexagonal boron nitride (hBN)/graphene
heterostructure. We assume the graphene layers are rotated symmetrically with
respect to the $A A'$-stacked hBN slab, with rotation angles $\theta_t =
\theta / 2$ for the top graphene and $\theta_b = - \theta / 2$ for the bottom
graphene. The coordinate system that we use is illustrated in Fig.~\ref{fig:stacking}.

Graphene is described by a Dirac Hamiltonian {\cite{graphene_Hamiltonian}}
\begin{equation}
  \hat{H}_{t (b) \nu \mathbf{k}} = \hbar v_{\tmop{Gr}} (\nu \sigma_x k_{x'} +
  \sigma_y k_{y'}) + U_{t (b)}, \label{eq:Gr_Hamiltonian}
\end{equation}
where $v_{\tmop{Gr}}$ is the Dirac velocity in graphene, $\sigma_x$,
$\sigma_y$ are the Pauli matrices, $\mathbf{k}$ is the electron momentum
measured from the nearest Dirac point, and $\nu = \pm 1$ is the valley index.
$k_{x'}, k_{y'}$ are the components of $\mathbf{k}$ in the reference frame
aligned with the corresponding graphene layer:
\begin{equation}
  \left(\begin{array}{c}
    k_{x'}\\
    k_{y'}
  \end{array}\right) = \left(\begin{array}{cc}
    \cos \theta_{t (b)} & \sin \theta_{t (b)}\\
    - \sin \theta_{t (b)} & \cos \theta_{t (b)}
  \end{array}\right) \left(\begin{array}{c}
    k_x\\
    k_y
  \end{array}\right) . \label{eq:rotated_k}
\end{equation}
A vertical electric field may be present between the graphene layers, which is
included in (\ref{eq:Gr_Hamiltonian}) as a potential energy $U_{t (b)}$.

hBN layers are also described by a Dirac Hamiltonian
{\cite{Gr-hBN_tunneling_model_2,Gr-hBN_tunneling_model_3}}
\begin{equation}
  \hat{H}_{d \nu \mathbf{k}, i i} = \left\{\begin{array}{l}
    \left(\begin{array}{cc}
      \epsilon_B & 0\\
      0 & \epsilon_N
    \end{array}\right), \quad i \tmop{odd}\\
    \left(\begin{array}{cc}
      \epsilon_N & 0\\
      0 & \epsilon_B
    \end{array}\right), \quad i \tmop{even}
  \end{array}\right. + U_i \label{eq:mBN_Hamiltonian}
\end{equation}
(for $i$th layer), where $\epsilon_B$ and $\epsilon_N$ are the energies of
$p_z$ orbitals of boron and nitrogen measured from the Dirac point of
graphene. $\mathbf{k}$ dependence is neglected because of the large bandgap.
Again, we include potential energy $U_i$ in the vertical electric field.

Hopping between layers $i$ and $i + 1$ in $A A'$-stacked hBN is described by
the interlayer hopping Hamiltonian
{\cite{Gr-hBN_tunneling_model_3,hBN_tight-binding}}
\begin{equation}
  \hat{H}_{d\mathbf{k}, i, i + 1} = \hat{H}_{d\mathbf{k}, i + 1, i} =
  \left(\begin{array}{cc}
    t_{\tmop{BN}} & 0\\
    0 & t_{\tmop{BN}}
  \end{array}\right) \label{eq:hBN_interlayer_Hamiltonian} .
\end{equation}

Hopping between the top (bottom) graphene layer and the top (bottom) hBN layer
is described by the following Hamiltonians
{\cite{Gr-hBN_tunneling_model_2,Gr-hBN_coupling,Gr-hBN_coupling_2,twisted_coupling}}:
\begin{eq}{Gr-hBN_coupling}
  \hat{T}_{t, \nu \tmmathbf{k}\tmmathbf{k}'}^{\dag} & = \frac{1}{3} \sum_{n
  = - 1}^1 (2 \pi)^2 \delta (\mathbf{K}_{t, \nu n}
  +\mathbf{k}-\mathbf{K}_{\tmop{hBN}, \nu n} -\mathbf{k}')\\ &\times
  \left(\begin{array}{cc}
    t_{\tmop{CB}} & t_{\tmop{CN}} e^{- \frac{2 \pi n i}{3}}\\
    t_{\tmop{CB}} e^{\frac{2 \pi n i}{3}} & t_{\tmop{CN}}
  \end{array}\right), \\
  \hat{T}_{b, \nu \tmmathbf{k}\tmmathbf{k}'}^{\dag} & = \frac{1}{3} \sum_{n
  = - 1}^1 (2 \pi)^2 \delta (\mathbf{K}_{b, \nu n}
  +\mathbf{k}-\mathbf{K}_{\tmop{hBN}, \nu n} -\mathbf{k}')\\ &\times
  \left\{\begin{array}{l}
    \left(\begin{array}{cc}
      t_{\tmop{CB}} & t_{\tmop{CN}} e^{- \frac{2 \pi n i}{3}}\\
      t_{\tmop{CB}} e^{\frac{2 \pi n i}{3}} & t_{\tmop{CN}}
    \end{array}\right), N \tmop{odd},\\
    \left(\begin{array}{cc}
      t_{\tmop{CN}} & t_{\tmop{CB}} e^{- \frac{2 \pi n i}{3}}\\
      t_{\tmop{CN}} e^{\frac{2 \pi n i}{3}} & t_{\tmop{CB}}
    \end{array}\right), N \tmop{even},
  \end{array}\right.
\end{eq}
where
\begin{eq}{Dirac_points_momenta}
  \mathbf{K}_{\tmop{hBN}, \nu n} & = \frac{4 \pi}{3 a_{\tmop{hBN}}}
  \left(\begin{array}{c}
    \nu \cos \frac{2 \pi n}{3}\\
    \sin \frac{2 \pi n}{3}
  \end{array}\right), \\
  \mathbf{K}_{t (b), \nu n} & = \frac{4 \pi}{3 a_{\tmop{Gr}}}
  \left(\begin{array}{c}
    \nu \cos \left( \frac{2 \pi n}{3} + \nu \theta_{t (b)} \right)\\
    \sin \left( \frac{2 \pi n}{3} + \nu \theta_{t (b)} \right)
  \end{array}\right)
\end{eq}
are the momenta corresponding to the $K$, $K'$ points in the Brillouin zone of
hBN and graphene layers.

From eq. (\ref{eq:Gr-hBN_coupling}) it is evident that tunneling conserves
lateral momentum, provided (i) all six corners of the Brillouin zone are
regarded as inequivalent, (ii) we always choose the same $n$ in both
$\hat{T}_{t, \nu \tmmathbf{k}\tmmathbf{k}'}^{\dag}$ and $\hat{T}_{b, \nu
\tmmathbf{k}\tmmathbf{k}'}$ when calculating the tunneling matrix element. The
first condition means that we should use $g = 12$ in eq.
(\ref{eq:broadened_golden_rule}) (twofold spin degeneracy and sixfold valley
degeneracy). The latter condition discards Umklapp processes,
which may produce additional peaks in the $I$-$V$ curves {\cite{Gr-hBN_tunneling_model_4,Gr-hBN_tunneling_model,Gr-hBN_tunneling_model_3}}, but are ignored in this work to simplify discussion.

Before we proceed with the calculations, we will clarify a few technical
details.

\begin{widefig}{I-V}{figs/I-V}
  $I$-$V$ curves of a graphene/hBN/graphene
  heterostructure at various couplings to contacts $\gamma$ and misalignment
  angles $\theta$. (a) Fixed Fermi energies $E_{F t}$ = 0, $E_{F b}$ = 0.2 eV;
  (b--c) fixed gate voltage $V_g$ in a single-gated heterostructure; (d) fixed
  gate voltages $V_{t g}, V_{b g}$ in a double-gated heterostructure. The
  tunnel barrier is 2 monolayers thick in (a--b) and 1 monolayer thick in
  (c--d). Insets in (a) show the band alignments corresponding to the
  resonance conditions with or without rotational misalignment. All the data
  is for 0 K.
\end{widefig}

First, we need to adapt the single-band equation
(\ref{eq:broadened_golden_rule}) to two-band graphene. The adaptation is
rather straightforward: we (i) insert the band indices $s_t$, $s_b$ where
appropriate, (ii) use graphene spinors $\psi^{\dag}_{s_t \nu \mathbf{k}_t}$,
$\psi_{s_b \nu \mathbf{k}_b}$ when calculating the tunneling matrix element
($v_{t b, s_t s_b \nu \mathbf{k}_t} = \psi^{\dag}_{s_t \nu \mathbf{k}_t}
\hat{T}_{t, \nu \tmmathbf{k}_t \tmmathbf{k}_d}^{\dag} \hat{g}^R_{d \nu
\tmmathbf{k}_d} \hat{T}_{b, \nu \tmmathbf{k}_d \tmmathbf{k}_b} \psi_{s_b \nu
\mathbf{k}_b}$), and (iii) sum over bands when calculating the total tunneling
current. This is justified because the tunnel coupling between electron states
is typically pairwise (at each momentum, tunneling occurs mostly between the
two bands that are the closest in energy).

Second, interpreting the results is easier with a constant $v_{t b}$, while it
actually depends on the momentum direction. Thus, we use $| v_{t b\mathbf{k}}
|^2$ averaged over the parts of the Brillouin zone where either
$\epsilon_{t\mathbf{k}}$ or $\epsilon_{b\mathbf{k}}$ lies between the Fermi
levels of the leads.

Third, if the twist angle $\theta$ is exactly zero, three possible tunneling
paths for each electron (different terms in eq. (\ref{eq:Gr-hBN_coupling}))
end in the same state and interfere with each other. We avoid this special
case by using an infinitesimal $\theta$ instead of $\theta = 0$ when dealing
with perfectly aligned heterostructures.

We use the following values for the model parameters: lattice constant of
graphene $a_{\tmop{Gr}} = 0.2461$ nm {\cite{Gr_lattice_constant}}, lattice
constant of hBN $a_{\tmop{hBN}} = 0.2505$ nm {\cite{hBN_lattice_constant}},
Dirac velocity in graphene $v_{\tmop{Gr}} = 10^6$ m/s, carbon-boron and
carbon-nitrogen hopping parameters $t_{\tmop{CB}} = 0.39$ eV, $t_{\tmop{CN}} =
0.26$ eV {\cite{Gr-hBN_coupling}}. For the on-site energies at the boron and
nitrogen atoms, we use values $\epsilon_B = - \epsilon_N = 3$ eV to match the
experimentally measured bandgap of $\sim$6--7 eV
{\cite{mBN_direct_gap,hBN_indirect_gap,hBN_gap_photocurrent,Gr-hBN_band_alignment,mBN_gap_ARPES,hBN_gap_R_vs_T,hBN_gap_R_vs_T_2,hBN_gap_R_vs_T_3}}
and band alignment with graphite
{\cite{Gr-hBN_band_alignment,Gr-hBN_band_alignment_2}}. Although the direct
electronic gap at the $K$ point of hBN (relevant for tunneling) is not as well
known as the usually measured optical gap, this uncertainty can be compensated
for by an appropriate choice of the interlayer hopping parameter
$t_{\tmop{BN}}$. We set $t_{\tmop{BN}} = 0.4$ eV to reproduce the experimental
dependence of the tunneling current on the number of hBN layers, $J \propto
\left( \frac{1}{50} \right)^{N_{\tmop{layers}}}$
{\cite{current_vs_Nlayers,current_vs_Nlayers_2}}.

%% file: results.tex
\section{Results and discussion}

We use eq. (\ref{eq:broadened_golden_rule}) to calculate the tunneling current
in a graphene/hBN (2 layers)/graphene/SiO$_2$ (300 nm)/gate heterostructure at
different couplings to contacts $\gamma$ and misalignment angles $\theta$.
Similar heterostructures, but with a 4-layer tunneling barrier were studied
experimentally in \Cite{Gr-hBN_tunneling_model_2}.

The calculated $I$-$V$ curves at fixed Fermi energies $E_{F t}$ = 0, $E_{F b}$
= 0.2 eV are shown in Fig.~\ref{fig:I-V}a. We can observe a resonance peak at
$V_{\tmop{bias}} = 0.2$ V, where the Dirac points of both graphenes lie at the
same energy.

With two-layer hBN as the tunneling barrier, the average tunneling matrix
element $| v_{t b} |$ is 1.2 meV. Therefore, the maximum possible tunneling
current is achieved at $\gamma = 2 | v_{t b} | = 2.4$ meV (green curve in Fig.~\ref{fig:I-V}a). 
At both smaller and larger $\gamma$, the amplitude of the
resonance peak is reduced.

The full width at half maximum of the resonance peak is $2 \sqrt{4 | v_{t b}
|^2 + \gamma^2} \approx 4 | v_{t b} | = 4.8$ mV at $\gamma \ll 2 | v_{t b} |$,
slightly increases to $4 \sqrt{2} | v_{t b} | = 6.8$ mV at $\gamma = 2 | v_{t
b} |$ (which maximizes the current), and continues to grow as $2 \gamma$ at
$\gamma \gg 2 | v_{t b} |$.

If the graphene layers are doped electrostatically, it is easier to fix the
gate voltage than the Fermi energies in individual layers. In this case, the
resonance width acquires an extra factor of $f_C = 1 + C_d / C_{q t} + C_d /
C_{q b}$, where $C_{q t}$, $C_{q b}$ are the quantum capacitances of the
graphene layers, and $C_d$ is the classical capacitance of the hBN tunneling
barrier {\cite{tunneling_Gr_coherence_length}}. In a single-gated
heterostructure, this factor is rather big because the top graphene remains
weakly doped and has a small quantum capacitance. This effect is illustrated
in Fig.~\ref{fig:I-V}b, where $I$-$V$ curves are calculated at a constant gate
voltage of 40 V. At $\gamma = 2.4$ meV (green curve), the peak width is 72 mV---vs 6.8 mV at fixed Fermi energies.

While coupling to the contacts and tunneling-induced level splitting present
fundamental limits to the sharpness of the tunneling resonance, in practice
there are other factors that contribute to the broadening and attenuation of
the resonance peak.

One of these factors is rotational misalignment between the top and bottom
graphene. Even a small misalignment of 0.1{\textdegree} is enough to cause a
threefold reduction in the maximum current (brown curves in Figs.~\ref{fig:I-V}a, 
\ref{fig:I-V}b).

Although it makes impossible to achieve perfect band alignment at any
$V_{\tmop{bias}}$, rotational misalignment does not {\tmem{per se}} eliminate
the singularity in tunneling current. Tunneling conductance becomes singular
not only when the Fermi circles in both leads coincide, but also when they
merely touch {\cite{tunneling_disorder_sqrt_divergence}}. Graphene's linear
dispersion has a peculiar property: when the Dirac points in the leads are
mismatched by $\Delta U$ in energy and by $\mathbf{q}$ in momentum, and
$\Delta U = \pm \hbar v_{\tmop{Gr}} q$, then the Fermi circles touch at
{\tmem{every}} energy (insets in Fig.~\ref{fig:I-V}a). The singularity in
conductance survives integration over energy and becomes a square-root
singularity in the tunneling current, $J \propto 1 / \sqrt{(\hbar
v_{\tmop{Gr}} q)^2 - \Delta U^2}$
{\cite{Gr_sqrt_divergence,Gr-hBN_tunneling_model_3}}.

Therefore, rotational misalignment of graphene leads merely splits the
resonant delta peak in the tunneling current into two square-root
singularities. To obtain a finite current at the ``skew resonances'' ($\Delta
U = \pm \hbar v_{\tmop{Gr}} q$), we still need to consider the effects of
finite $\gamma$ and $v_{t b}$.

Using eq. (\ref{eq:convolved_current}), one can derive the following
approximate shape of the resonant peaks in the presence of rotational
misalignment:
\begin{eq}{broadened_sqrt}
  J \propto \frac{\gamma}{\gamma_{\tmop{eff}}} \tmop{Re}
  \frac{V_{\tmop{bias}}}{\sqrt{(\hbar v_{\tmop{Gr}} q)^2 - \left(
  \frac{V_{\tmop{bias}} - V_{\tmop{res}}}{f_C} + i \gamma_{\tmop{eff}} \right)^2}},
\end{eq}
where $\gamma_{\tmop{eff}} = \sqrt{\gamma^2 + 4 | v_{t b} |^2}$, $V_{\tmop{res}}$ 
is the resonance position at $\theta = 0$, and
we have inserted an extra factor of $V_{\tmop{bias}}$ to get linear
behavior at $V_{\tmop{bias}} \rightarrow 0$.

The dependence of the peak height and width on $\gamma$ and $v_{t b}$ (dashed
curves in Fig.~\ref{fig:I-V}d) is similar to the rotationally aligned case
(solid curves in Fig.~\ref{fig:I-V}d). Of course, the maximum conductance is
smaller than in the rotationally aligned case (when {\tmem{all}} electrons
with energies between $\mu_t$, $\mu_b$ participate in tunneling), but an
increase in the rotational misalignment does not always reduce the maximum
current (Fig.~\ref{fig:I-V}c), although the resonant peak may be shifted
outside the experimentally accessible region. This is because the reduction in
conductance can be compensated by the increase in the resonant bias voltage,
which is significant at large misalignment angles (Fig.~\ref{fig:I-V}c).

Another factor contributing to the resonance broadening is electron
scattering. Including scattering in calculations is a standard way of
obtaining a finite current at the tunneling resonance
{\cite{Gr_tunneling_theory_review}}. There is a number of papers about the
effects of static disorder
{\cite{tunneling_disorder_sqrt_divergence,tunneling_Gr_coherence_length,tunneling_interface_roughness,tunneling_Gr_correlated_disorder_divergence,Gr-hBN_tunneling_model_3}},
electron-phonon (e-ph) {\cite{Gr-hBN_tunneling_model_3,Vdovin_e-ph}}, electron-electron (e-e)
{\cite{tunneling_e-e,tunneling_e-e_e-pl,Gr-hBN_tunneling_model_e-e}}, and
electron-plasmon (e-pl) scattering {\cite{tunneling_e-e_e-pl}}
on tunneling characteristics.

Scattering affects the resonant peak in much the same way as coupling to the
contacts: compare eq. (\ref{eq:resonance_height}) of this paper with 
eq. (11) of \Cite{tunneling_disorder_sqrt_divergence},
eq. (6) of \Cite{tunneling_e-e},
eq. (9) of \Cite{tunneling_e-e_e-pl},
eq. (29) of \Cite{tunneling_interface_roughness}. 
To achieve the sharpest possible resonance, the
scattering rate $\gamma_{\tmop{sc}}$ should be less than the tunnel level
splitting $2 | v_{t b} |$.

Advances in graphene technology make it possible to achieve sub-meV elastic
scattering rates {\cite{Gr_record_mean_free_path}} (although one should be
careful when estimating the quantum lifetime from transport measurements
{\cite{transport_vs_quasiparticle_relaxation_time, Gr_quantum_lifetime_vs_substrate}}).

Inelastic processes, such as e-e and e-ph scattering, are more difficult to
suppress than elastic scattering. Since the resonant peak is located at a
finite $V_{\tmop{bias}}$, the system is in a nonequilibrium state and can have
a significant inelastic scattering rate even at zero temperature. The e-e
scattering rate can be estimated as $\gamma_{e - e} \approx \frac{(e
V_{\tmop{bias}})^2}{4 \pi E_F} \left[ \ln \left( \frac{8 E_F}{e
V_{\tmop{bias}}} \right) - \frac{1}{2} \right]$ {\cite{Gr_e-e_e-ph_rates}}
$\approx$ a few meV or tens of meV for typical values of $V_{\tmop{bias}}$ and
$E_F$ \cite{Gr-hBN_SiO2_quantum_lifetime, Gr-hBN_quantum_lifetime, BLG-WSe2_quantum_lifetime, BLG_Gr_quantum_lifetime_ab_initio}. 
The e-ph scattering rate is typically in the sub-meV -- few meV range 
{\cite{Gr_e-e_e-ph_rates,Gr_e-ph_rate}}.

Therefore, to observe the effect considered in this paper (when the height and
width of the tunneling resonance are limited by tunneling itself), a number of
conditions must be met: (i) High-quality graphene should be used to minimize
the elastic scattering rate. (ii) The resonant peak should be located at small
$V_{\tmop{bias}}$. In this case, the tunneling electrons are close to the
Fermi levels of both graphenes, and inelastic scattering is suppressed. (iii)
The tunnel barrier must be very thin so that $2 | v_{t b} |$ exceeds the
scattering rate ($\sim$ a few meV). (iv) Both graphenes must be strongly doped
(have a large quantum capacitance) to ensure effective control of band
alignment by the bias voltage (i. e., to minimize the $f_C$ factor discussed
above). Large Fermi energies also suppress inelastic scattering. (v) The
coupling to contacts $\gamma$ should be close to $2 | v_{t b} |$.

Condition (i) favors graphene encapsulated in hBN, which can demostrate
exceptional mobility {\cite{Gr_record_mean_free_path}}. Conditions (ii) and
(iv) require double gating, which provides independent control over the Fermi
energies of both graphenes. Condition (iii) is most easily satisfied with
single-layer hBN as the tunnel barrier. Thus, the optimal conditions for
observing a resonance whose width is limited by tunnel splitting are realized
in a top gate/hBN/graphene/hBN monolayer/graphene/hBN/bottom gate
heterostructure.

The calculated $I$-$V$ curves for such a device are presented in Fig.~\ref{fig:I-V}d. 
The peaks at $\theta = 0$ are significantly narrower than in a
similar single-gated heterostructure (Fig.~\ref{fig:I-V}c) because of the
higher quantum capacitance of the top graphene. However, they are still wider
than $2 \sqrt{4 | v_{t b} |^2 + \gamma^2}$ because the ultrathin barrier has a
capacitance comparable to the quantum capacitance of graphene even at large
Fermi energies. For example, at $\gamma = 2 | v_{t b} | = 10$ meV the peak
width is 69 mV instead of $2 \sqrt{4 | v_{t b} |^2 + \gamma^2} = 28$ mV.

If the misalignment between graphene and hBN is below 1{\textdegree}, an
incommensurate-commensurate transition may occur, when graphene lattice
spontaneously adapts to hBN lattice and domain walls form at every 14 nm
{\cite{Gr_commensurate-incommensurate}}. Note that it does not preclude
fabrication of high-quality heterostructures without domain walls where both
graphene layers are aligned with each other, but not with hBN. The hBN
orientation may affect the barrier height, but the main results of this paper
should remain unchanged. In the model we use in this paper, when hBN bands are
assumed dispersionless in the lateral directions and Umklapp processes are
neglected, the tunneling characteristics do not depend on the hBN orientation
at all.

To conclude, we have discussed the effect of the tunneling-induced level
shifts and the coupling to contacts on resonant tunneling in van der Waals
heterostructures, using a graphene/hBN/graphene heterostructure as an example.
We demonstrated that the resonant peak width cannot be less than $4 | v_{t b}
|$, where $v_{t b}$ is the tunneling matrix element, and the peak height is
maximal when the coupling to contacts $\gamma$ equals $2 | v_{t b} |$. We have
shown that these effects can be observed in high-quality double-gated
top gate/hBN/graphene/hBN/graphene/hBN/bottom gate heterostructures with ultrathin (1 or
2 monolayers) barriers, when the tunnel coupling exceeds the scattering rate.

%% file: acknowledgements.tex
\begin{acknowledgements}
This paper is written mostly in GNU TeXmacs~\cite{TeXmacs}, a convenient tool for mathematical typesetting.

The work was supported by the grant \# 21-79-20225 of the Russian Scientific Foundation.
\end{acknowledgements}

%% file: main.bbl
\begin{thebibliography}{71}%
\makeatletter
\providecommand \@ifxundefined [1]{%
 \@ifx{#1\undefined}
}%
\providecommand \@ifnum [1]{%
 \ifnum #1\expandafter \@firstoftwo
 \else \expandafter \@secondoftwo
 \fi
}%
\providecommand \@ifx [1]{%
 \ifx #1\expandafter \@firstoftwo
 \else \expandafter \@secondoftwo
 \fi
}%
\providecommand \natexlab [1]{#1}%
\providecommand \enquote  [1]{``#1''}%
\providecommand \bibnamefont  [1]{#1}%
\providecommand \bibfnamefont [1]{#1}%
\providecommand \citenamefont [1]{#1}%
\providecommand \href@noop [0]{\@secondoftwo}%
\providecommand \href [0]{\begingroup \@sanitize@url \@href}%
\providecommand \@href[1]{\@@startlink{#1}\@@href}%
\providecommand \@@href[1]{\endgroup#1\@@endlink}%
\providecommand \@sanitize@url [0]{\catcode `\\12\catcode `\$12\catcode `\&12\catcode `\#12\catcode `\^12\catcode `\_12\catcode `\%12\relax}%
\providecommand \@@startlink[1]{}%
\providecommand \@@endlink[0]{}%
\providecommand \url  [0]{\begingroup\@sanitize@url \@url }%
\providecommand \@url [1]{\endgroup\@href {#1}{\urlprefix }}%
\providecommand \urlprefix  [0]{URL }%
\providecommand \Eprint [0]{\href }%
\providecommand \doibase [0]{https://doi.org/}%
\providecommand \selectlanguage [0]{\@gobble}%
\providecommand \bibinfo  [0]{\@secondoftwo}%
\providecommand \bibfield  [0]{\@secondoftwo}%
\providecommand \translation [1]{[#1]}%
\providecommand \BibitemOpen [0]{}%
\providecommand \bibitemStop [0]{}%
\providecommand \bibitemNoStop [0]{.\EOS\space}%
\providecommand \EOS [0]{\spacefactor3000\relax}%
\providecommand \BibitemShut  [1]{\csname bibitem#1\endcsname}%
\let\auto@bib@innerbib\@empty
\bibitem [{\citenamefont {Schneider}\ and\ \citenamefont {Liu}(2007)}]{QWIPs}%
  \BibitemOpen
  \bibfield  {author} {\bibinfo {author} {\bibfnamefont {H.}~\bibnamefont {Schneider}}\ and\ \bibinfo {author} {\bibfnamefont {H.~C.}\ \bibnamefont {Liu}},\ }\href {https://books.google.com/books?id=f7a5BQAAQBAJ&pg=PR9} {\emph {\bibinfo {title} {Quantum well infrared photodetectors}}}\ (\bibinfo  {publisher} {Springer},\ \bibinfo {year} {2007})\BibitemShut {NoStop}%
\bibitem [{\citenamefont {Faist}\ \emph {et~al.}(1994)\citenamefont {Faist}, \citenamefont {Capasso}, \citenamefont {Sivco}, \citenamefont {Sirtori}, \citenamefont {Hutchinson},\ and\ \citenamefont {Cho}}]{QCL}%
  \BibitemOpen
  \bibfield  {author} {\bibinfo {author} {\bibfnamefont {J.}~\bibnamefont {Faist}}, \bibinfo {author} {\bibfnamefont {F.}~\bibnamefont {Capasso}}, \bibinfo {author} {\bibfnamefont {D.~L.}\ \bibnamefont {Sivco}}, \bibinfo {author} {\bibfnamefont {C.}~\bibnamefont {Sirtori}}, \bibinfo {author} {\bibfnamefont {A.~L.}\ \bibnamefont {Hutchinson}},\ and\ \bibinfo {author} {\bibfnamefont {A.~Y.}\ \bibnamefont {Cho}},\ }\bibfield  {title} {\bibinfo {title} {Quantum cascade laser},\ }\href {https://www.researchgate.net/profile/Jerome-Faist/publication/272026462_Quantum_Cascade_Laser/links/09e41513e03a5e53c4000000/Quantum-Cascade-Laser.pdf\#page=2} {\bibfield  {journal} {\bibinfo  {journal} {Science}\ }\textbf {\bibinfo {volume} {264}},\ \bibinfo {pages} {553} (\bibinfo {year} {1994})}\BibitemShut {NoStop}%
\bibitem [{\citenamefont {Wang}\ \emph {et~al.}(2019)\citenamefont {Wang}, \citenamefont {Yu}, \citenamefont {Lei}, \citenamefont {Zhu}, \citenamefont {Cao}, \citenamefont {Liu}, \citenamefont {You}, \citenamefont {Zeng}, \citenamefont {Deng}, \citenamefont {Zhu}, \citenamefont {Zhou}, \citenamefont {Fu}, \citenamefont {Wang}, \citenamefont {Huang},\ and\ \citenamefont {Liu}}]{Wang-vdW_TFET}%
  \BibitemOpen
  \bibfield  {author} {\bibinfo {author} {\bibfnamefont {X.}~\bibnamefont {Wang}}, \bibinfo {author} {\bibfnamefont {P.}~\bibnamefont {Yu}}, \bibinfo {author} {\bibfnamefont {Z.}~\bibnamefont {Lei}}, \bibinfo {author} {\bibfnamefont {C.}~\bibnamefont {Zhu}}, \bibinfo {author} {\bibfnamefont {X.}~\bibnamefont {Cao}}, \bibinfo {author} {\bibfnamefont {F.}~\bibnamefont {Liu}}, \bibinfo {author} {\bibfnamefont {L.}~\bibnamefont {You}}, \bibinfo {author} {\bibfnamefont {Q.}~\bibnamefont {Zeng}}, \bibinfo {author} {\bibfnamefont {Y.}~\bibnamefont {Deng}}, \bibinfo {author} {\bibfnamefont {C.}~\bibnamefont {Zhu}}, \bibinfo {author} {\bibfnamefont {J.}~\bibnamefont {Zhou}}, \bibinfo {author} {\bibfnamefont {Q.}~\bibnamefont {Fu}}, \bibinfo {author} {\bibfnamefont {J.}~\bibnamefont {Wang}}, \bibinfo {author} {\bibfnamefont {Y.}~\bibnamefont {Huang}},\ and\ \bibinfo {author} {\bibfnamefont {Z.}~\bibnamefont {Liu}},\ }\bibfield  {title} {\bibinfo {title} {{Van der Waals negative capacitance transistors}},\ }\href
  {https://www.nature.com/articles/s41467-019-10738-4.pdf} {\bibfield  {journal} {\bibinfo  {journal} {Nature Communications}\ }\textbf {\bibinfo {volume} {10}},\ \bibinfo {pages} {3037} (\bibinfo {year} {2019})}\BibitemShut {NoStop}%
\bibitem [{\citenamefont {Xiong}\ \emph {et~al.}(2020)\citenamefont {Xiong}, \citenamefont {Huang}, \citenamefont {Hu}, \citenamefont {Li}, \citenamefont {Liu}, \citenamefont {Li}, \citenamefont {Tian}, \citenamefont {Li}, \citenamefont {Song},\ and\ \citenamefont {Wu}}]{Xiong-vdW_TFET}%
  \BibitemOpen
  \bibfield  {author} {\bibinfo {author} {\bibfnamefont {X.}~\bibnamefont {Xiong}}, \bibinfo {author} {\bibfnamefont {M.}~\bibnamefont {Huang}}, \bibinfo {author} {\bibfnamefont {B.}~\bibnamefont {Hu}}, \bibinfo {author} {\bibfnamefont {X.}~\bibnamefont {Li}}, \bibinfo {author} {\bibfnamefont {F.}~\bibnamefont {Liu}}, \bibinfo {author} {\bibfnamefont {S.}~\bibnamefont {Li}}, \bibinfo {author} {\bibfnamefont {M.}~\bibnamefont {Tian}}, \bibinfo {author} {\bibfnamefont {T.}~\bibnamefont {Li}}, \bibinfo {author} {\bibfnamefont {J.}~\bibnamefont {Song}},\ and\ \bibinfo {author} {\bibfnamefont {Y.}~\bibnamefont {Wu}},\ }\bibfield  {title} {\bibinfo {title} {{A transverse tunnelling field-effect transistor made from a van der Waals heterostructure}},\ }\href
  {https://www.researchgate.net/profile/Mengchuan-Tian/publication/339010818_A_transverse_tunnelling_field-effect_transistor_made_from_a_van_der_Waals_heterostructure/links/5e53a0b9458515072db7aa8c/A-transverse-tunnelling-field-effect-transistor-made-from-a-van-der-Waals-heterostructure.pdf\#page=2} {\bibfield  {journal} {\bibinfo  {journal} {Nature Electronics}\ }\textbf {\bibinfo {volume} {3}},\ \bibinfo {pages} {106} (\bibinfo {year} {2020})}\BibitemShut {NoStop}%
\bibitem [{\citenamefont {Murphy}\ \emph {et~al.}(1995)\citenamefont {Murphy}, \citenamefont {Eisenstein}, \citenamefont {Pfeiffer},\ and\ \citenamefont {West}}]{lifetime_by_tun_spectrocopy_Murphy}%
  \BibitemOpen
  \bibfield  {author} {\bibinfo {author} {\bibfnamefont {S.~Q.}\ \bibnamefont {Murphy}}, \bibinfo {author} {\bibfnamefont {J.~P.}\ \bibnamefont {Eisenstein}}, \bibinfo {author} {\bibfnamefont {L.~N.}\ \bibnamefont {Pfeiffer}},\ and\ \bibinfo {author} {\bibfnamefont {K.~W.}\ \bibnamefont {West}},\ }\bibfield  {title} {\bibinfo {title} {Lifetime of two-dimensional electrons measured by tunneling spectroscopy},\ }\href {https://arxiv.org/pdf/cond-mat/9506042} {\bibfield  {journal} {\bibinfo  {journal} {Phys. Rev. B}\ }\textbf {\bibinfo {volume} {52}},\ \bibinfo {pages} {14825} (\bibinfo {year} {1995})}\BibitemShut {NoStop}%
\bibitem [{\citenamefont {Turner}\ \emph {et~al.}(1996)\citenamefont {Turner}, \citenamefont {Nicholls}, \citenamefont {Linfield}, \citenamefont {Brown}, \citenamefont {Jones},\ and\ \citenamefont {Ritchie}}]{tunnel_spectroscopy_Turner}%
  \BibitemOpen
  \bibfield  {author} {\bibinfo {author} {\bibfnamefont {N.}~\bibnamefont {Turner}}, \bibinfo {author} {\bibfnamefont {J.~T.}\ \bibnamefont {Nicholls}}, \bibinfo {author} {\bibfnamefont {E.~H.}\ \bibnamefont {Linfield}}, \bibinfo {author} {\bibfnamefont {K.~M.}\ \bibnamefont {Brown}}, \bibinfo {author} {\bibfnamefont {G.~A.~C.}\ \bibnamefont {Jones}},\ and\ \bibinfo {author} {\bibfnamefont {D.~A.}\ \bibnamefont {Ritchie}},\ }\bibfield  {title} {\bibinfo {title} {Tunneling between parallel two-dimensional electron gases},\ }\href {https://arxiv.org/pdf/cond-mat/9609067\#page=24} {\bibfield  {journal} {\bibinfo  {journal} {Phys. Rev. B}\ }\textbf {\bibinfo {volume} {54}},\ \bibinfo {pages} {10614} (\bibinfo {year} {1996})}\BibitemShut {NoStop}%
\bibitem [{\citenamefont {Jang}\ \emph {et~al.}(2017)\citenamefont {Jang}, \citenamefont {Yoo}, \citenamefont {Pfeiffer}, \citenamefont {West}, \citenamefont {Baldwin},\ and\ \citenamefont {Ashoori}}]{Spectral_function_resolved_by_tunneling}%
  \BibitemOpen
  \bibfield  {author} {\bibinfo {author} {\bibfnamefont {J.}~\bibnamefont {Jang}}, \bibinfo {author} {\bibfnamefont {H.~M.}\ \bibnamefont {Yoo}}, \bibinfo {author} {\bibfnamefont {L.~N.}\ \bibnamefont {Pfeiffer}}, \bibinfo {author} {\bibfnamefont {K.~W.}\ \bibnamefont {West}}, \bibinfo {author} {\bibfnamefont {K.~W.}\ \bibnamefont {Baldwin}},\ and\ \bibinfo {author} {\bibfnamefont {R.~C.}\ \bibnamefont {Ashoori}},\ }\bibfield  {title} {\bibinfo {title} {Full momentum- and energy-resolved spectral function of a 2d electronic system},\ }\href {https://arxiv.org/pdf/1701.01684\#page=3} {\bibfield  {journal} {\bibinfo  {journal} {Science}\ }\textbf {\bibinfo {volume} {358}},\ \bibinfo {pages} {901} (\bibinfo {year} {2017})}\BibitemShut {NoStop}%
\bibitem [{\citenamefont {Prasad}\ \emph {et~al.}(2021)\citenamefont {Prasad}, \citenamefont {Burg}, \citenamefont {Watanabe}, \citenamefont {Taniguchi}, \citenamefont {Register},\ and\ \citenamefont {Tutuc}}]{BLG-WSe2_quantum_lifetime}%
  \BibitemOpen
  \bibfield  {author} {\bibinfo {author} {\bibfnamefont {N.}~\bibnamefont {Prasad}}, \bibinfo {author} {\bibfnamefont {G.~W.}\ \bibnamefont {Burg}}, \bibinfo {author} {\bibfnamefont {K.}~\bibnamefont {Watanabe}}, \bibinfo {author} {\bibfnamefont {T.}~\bibnamefont {Taniguchi}}, \bibinfo {author} {\bibfnamefont {L.~F.}\ \bibnamefont {Register}},\ and\ \bibinfo {author} {\bibfnamefont {E.}~\bibnamefont {Tutuc}},\ }\bibfield  {title} {\bibinfo {title} {Quantum lifetime spectroscopy and magnetotunneling in double bilayer graphene heterostructures},\ }\href {https://arxiv.org/pdf/2105.07104.pdf\#page=3} {\bibfield  {journal} {\bibinfo  {journal} {Phys. Rev. Lett.}\ }\textbf {\bibinfo {volume} {127}},\ \bibinfo {pages} {117701} (\bibinfo {year} {2021})}\BibitemShut {NoStop}%
\bibitem [{\citenamefont {Spielman}\ \emph {et~al.}(2000)\citenamefont {Spielman}, \citenamefont {Eisenstein}, \citenamefont {Pfeiffer},\ and\ \citenamefont {West}}]{Strong_correlations_tunneling}%
  \BibitemOpen
  \bibfield  {author} {\bibinfo {author} {\bibfnamefont {I.~B.}\ \bibnamefont {Spielman}}, \bibinfo {author} {\bibfnamefont {J.~P.}\ \bibnamefont {Eisenstein}}, \bibinfo {author} {\bibfnamefont {L.~N.}\ \bibnamefont {Pfeiffer}},\ and\ \bibinfo {author} {\bibfnamefont {K.~W.}\ \bibnamefont {West}},\ }\bibfield  {title} {\bibinfo {title} {Resonantly enhanced tunneling in a double layer quantum {H}all ferromagnet},\ }\href {https://scholar.archive.org/work/yh7nubsirfacrgbesx3lcfm5bq/access/wayback/http://www.its.caltech.edu/~jpelab/papers/PRL84_5808_2000.pdf} {\bibfield  {journal} {\bibinfo  {journal} {Phys. Rev. Lett.}\ }\textbf {\bibinfo {volume} {84}},\ \bibinfo {pages} {5808} (\bibinfo {year} {2000})}\BibitemShut {NoStop}%
\bibitem [{\citenamefont {Nandi}\ \emph {et~al.}(2013)\citenamefont {Nandi}, \citenamefont {Khaire}, \citenamefont {Finck}, \citenamefont {Eisenstein}, \citenamefont {Pfeiffer},\ and\ \citenamefont {West}}]{Tunneling_in_quantum_Hall_bilayers}%
  \BibitemOpen
  \bibfield  {author} {\bibinfo {author} {\bibfnamefont {D.}~\bibnamefont {Nandi}}, \bibinfo {author} {\bibfnamefont {T.}~\bibnamefont {Khaire}}, \bibinfo {author} {\bibfnamefont {A.~D.~K.}\ \bibnamefont {Finck}}, \bibinfo {author} {\bibfnamefont {J.~P.}\ \bibnamefont {Eisenstein}}, \bibinfo {author} {\bibfnamefont {L.~N.}\ \bibnamefont {Pfeiffer}},\ and\ \bibinfo {author} {\bibfnamefont {K.~W.}\ \bibnamefont {West}},\ }\bibfield  {title} {\bibinfo {title} {Tunneling at ${\ensuremath{\nu}}_{T}=1$ in quantum {H}all bilayers},\ }\href {https://arxiv.org/pdf/1308.6269} {\bibfield  {journal} {\bibinfo  {journal} {Phys. Rev. B}\ }\textbf {\bibinfo {volume} {88}},\ \bibinfo {pages} {165308} (\bibinfo {year} {2013})}\BibitemShut {NoStop}%
\bibitem [{\citenamefont {Mishchenko}\ \emph {et~al.}(2014)\citenamefont {Mishchenko}, \citenamefont {Tu}, \citenamefont {Cao}, \citenamefont {Gorbachev}, \citenamefont {Wallbank}, \citenamefont {Greenaway}, \citenamefont {Morozov}, \citenamefont {Morozov}, \citenamefont {Zhu}, \citenamefont {Wong}, \citenamefont {Withers}, \citenamefont {Woods}, \citenamefont {Kim}, \citenamefont {Watanabe}, \citenamefont {Taniguchi}, \citenamefont {Vdovin}, \citenamefont {Makarovsky}, \citenamefont {Fromhold}, \citenamefont {Fal'ko}, \citenamefont {Geim}, \citenamefont {Eaves},\ and\ \citenamefont {Novoselov}}]{Gr-hBN_tunneling_model_2}%
  \BibitemOpen
  \bibfield  {author} {\bibinfo {author} {\bibfnamefont {A.}~\bibnamefont {Mishchenko}}, \bibinfo {author} {\bibfnamefont {J.}~\bibnamefont {Tu}}, \bibinfo {author} {\bibfnamefont {Y.}~\bibnamefont {Cao}}, \bibinfo {author} {\bibfnamefont {R.~V.}\ \bibnamefont {Gorbachev}}, \bibinfo {author} {\bibfnamefont {J.}~\bibnamefont {Wallbank}}, \bibinfo {author} {\bibfnamefont {M.}~\bibnamefont {Greenaway}}, \bibinfo {author} {\bibfnamefont {V.}~\bibnamefont {Morozov}}, \bibinfo {author} {\bibfnamefont {S.}~\bibnamefont {Morozov}}, \bibinfo {author} {\bibfnamefont {M.}~\bibnamefont {Zhu}}, \bibinfo {author} {\bibfnamefont {S.}~\bibnamefont {Wong}}, \bibinfo {author} {\bibfnamefont {F.}~\bibnamefont {Withers}}, \bibinfo {author} {\bibfnamefont {C.~R.}\ \bibnamefont {Woods}}, \bibinfo {author} {\bibfnamefont {Y.-J.}\ \bibnamefont {Kim}}, \bibinfo {author} {\bibfnamefont {K.}~\bibnamefont {Watanabe}}, \bibinfo {author} {\bibfnamefont {T.}~\bibnamefont {Taniguchi}}, \bibinfo {author} {\bibfnamefont {E.~E.}\ \bibnamefont
  {Vdovin}}, \bibinfo {author} {\bibfnamefont {O.}~\bibnamefont {Makarovsky}}, \bibinfo {author} {\bibfnamefont {T.~M.}\ \bibnamefont {Fromhold}}, \bibinfo {author} {\bibfnamefont {V.~I.}\ \bibnamefont {Fal'ko}}, \bibinfo {author} {\bibfnamefont {A.~K.}\ \bibnamefont {Geim}}, \bibinfo {author} {\bibfnamefont {L.}~\bibnamefont {Eaves}},\ and\ \bibinfo {author} {\bibfnamefont {K.~S.}\ \bibnamefont {Novoselov}},\ }\bibfield  {title} {\bibinfo {title} {Twist-controlled resonant tunnelling in graphene/boron nitride/graphene heterostructures},\ }\href {https://static-content.springer.com/esm/art\%3A10.1038\%2Fnnano.2014.187/MediaObjects/41565\_2014\_BFnnano2014187\_MOESM6\_ESM.pdf\#page=1} {\bibfield  {journal} {\bibinfo  {journal} {Nature nanotechnology}\ }\textbf {\bibinfo {volume} {9}},\ \bibinfo {pages} {808} (\bibinfo {year} {2014})}\BibitemShut {NoStop}%
\bibitem [{\citenamefont {Vdovin}\ \emph {et~al.}(2016)\citenamefont {Vdovin}, \citenamefont {Mishchenko}, \citenamefont {Greenaway}, \citenamefont {Zhu}, \citenamefont {Ghazaryan}, \citenamefont {Misra}, \citenamefont {Cao}, \citenamefont {Morozov}, \citenamefont {Makarovsky}, \citenamefont {Fromhold}, \citenamefont {Patan\`e}, \citenamefont {Slotman}, \citenamefont {Katsnelson}, \citenamefont {Geim}, \citenamefont {Novoselov},\ and\ \citenamefont {Eaves}}]{Vdovin_e-ph}%
  \BibitemOpen
  \bibfield  {author} {\bibinfo {author} {\bibfnamefont {E.~E.}\ \bibnamefont {Vdovin}}, \bibinfo {author} {\bibfnamefont {A.}~\bibnamefont {Mishchenko}}, \bibinfo {author} {\bibfnamefont {M.~T.}\ \bibnamefont {Greenaway}}, \bibinfo {author} {\bibfnamefont {M.~J.}\ \bibnamefont {Zhu}}, \bibinfo {author} {\bibfnamefont {D.}~\bibnamefont {Ghazaryan}}, \bibinfo {author} {\bibfnamefont {A.}~\bibnamefont {Misra}}, \bibinfo {author} {\bibfnamefont {Y.}~\bibnamefont {Cao}}, \bibinfo {author} {\bibfnamefont {S.~V.}\ \bibnamefont {Morozov}}, \bibinfo {author} {\bibfnamefont {O.}~\bibnamefont {Makarovsky}}, \bibinfo {author} {\bibfnamefont {T.~M.}\ \bibnamefont {Fromhold}}, \bibinfo {author} {\bibfnamefont {A.}~\bibnamefont {Patan\`e}}, \bibinfo {author} {\bibfnamefont {G.~J.}\ \bibnamefont {Slotman}}, \bibinfo {author} {\bibfnamefont {M.~I.}\ \bibnamefont {Katsnelson}}, \bibinfo {author} {\bibfnamefont {A.~K.}\ \bibnamefont {Geim}}, \bibinfo {author} {\bibfnamefont {K.~S.}\ \bibnamefont {Novoselov}},\ and\ \bibinfo
  {author} {\bibfnamefont {L.}~\bibnamefont {Eaves}},\ }\bibfield  {title} {\bibinfo {title} {Phonon-assisted resonant tunneling of electrons in graphene--boron nitride transistors},\ }\href {https://arxiv.org/pdf/1512.02143.pdf\#page=3} {\bibfield  {journal} {\bibinfo  {journal} {Phys. Rev. Lett.}\ }\textbf {\bibinfo {volume} {116}},\ \bibinfo {pages} {186603} (\bibinfo {year} {2016})}\BibitemShut {NoStop}%
\bibitem [{\citenamefont {Greenaway}\ \emph {et~al.}(2018)\citenamefont {Greenaway}, \citenamefont {Vdovin}, \citenamefont {Ghazaryan}, \citenamefont {Misra}, \citenamefont {Mishchenko}, \citenamefont {Cao}, \citenamefont {Wang}, \citenamefont {Wallbank}, \citenamefont {Holwill}, \citenamefont {Khanin}, \citenamefont {Morozov}, \citenamefont {Watanabe}, \citenamefont {Taniguchi}, \citenamefont {Makarovsky}, \citenamefont {Fromhold}, \citenamefont {Patan{\`{e}}}, \citenamefont {Geim}, \citenamefont {Fal'ko}, \citenamefont {Novoselov},\ and\ \citenamefont {Eaves}}]{Greenaway_Tunnel_spectroscopy}%
  \BibitemOpen
  \bibfield  {author} {\bibinfo {author} {\bibfnamefont {M.~T.}\ \bibnamefont {Greenaway}}, \bibinfo {author} {\bibfnamefont {E.~E.}\ \bibnamefont {Vdovin}}, \bibinfo {author} {\bibfnamefont {D.}~\bibnamefont {Ghazaryan}}, \bibinfo {author} {\bibfnamefont {A.}~\bibnamefont {Misra}}, \bibinfo {author} {\bibfnamefont {A.}~\bibnamefont {Mishchenko}}, \bibinfo {author} {\bibfnamefont {Y.}~\bibnamefont {Cao}}, \bibinfo {author} {\bibfnamefont {Z.}~\bibnamefont {Wang}}, \bibinfo {author} {\bibfnamefont {J.~R.}\ \bibnamefont {Wallbank}}, \bibinfo {author} {\bibfnamefont {M.}~\bibnamefont {Holwill}}, \bibinfo {author} {\bibfnamefont {Y.}~\bibnamefont {Khanin}}, \bibinfo {author} {\bibfnamefont {S.~V.}\ \bibnamefont {Morozov}}, \bibinfo {author} {\bibfnamefont {K.}~\bibnamefont {Watanabe}}, \bibinfo {author} {\bibfnamefont {T.}~\bibnamefont {Taniguchi}}, \bibinfo {author} {\bibfnamefont {O.}~\bibnamefont {Makarovsky}}, \bibinfo {author} {\bibfnamefont {T.~M.}\ \bibnamefont {Fromhold}}, \bibinfo {author} {\bibfnamefont
  {A.}~\bibnamefont {Patan{\`{e}}}}, \bibinfo {author} {\bibfnamefont {A.~K.}\ \bibnamefont {Geim}}, \bibinfo {author} {\bibfnamefont {V.~I.}\ \bibnamefont {Fal'ko}}, \bibinfo {author} {\bibfnamefont {K.~S.}\ \bibnamefont {Novoselov}},\ and\ \bibinfo {author} {\bibfnamefont {L.}~\bibnamefont {Eaves}},\ }\bibfield  {title} {\bibinfo {title} {Tunnel spectroscopy of localised electronic states in hexagonal boron nitride},\ }\href {https://arxiv.org/pdf/1810.01312.pdf\#page=3} {\bibfield  {journal} {\bibinfo  {journal} {Communications Physics}\ }\textbf {\bibinfo {volume} {1}},\ \bibinfo {pages} {94} (\bibinfo {year} {2018})}\BibitemShut {NoStop}%
\bibitem [{\citenamefont {Wallbank}\ \emph {et~al.}(2016)\citenamefont {Wallbank}, \citenamefont {Ghazaryan}, \citenamefont {Misra}, \citenamefont {Cao}, \citenamefont {Tu}, \citenamefont {Piot}, \citenamefont {Potemski}, \citenamefont {Pezzini}, \citenamefont {Wiedmann}, \citenamefont {Zeitler}, \citenamefont {Lane}, \citenamefont {Morozov}, \citenamefont {Greenaway}, \citenamefont {Eaves}, \citenamefont {Geim}, \citenamefont {Fal'ko}, \citenamefont {Novoselov},\ and\ \citenamefont {Mishchenko}}]{Wallbank_tuning_valley_with_tunneling}%
  \BibitemOpen
  \bibfield  {author} {\bibinfo {author} {\bibfnamefont {J.~R.}\ \bibnamefont {Wallbank}}, \bibinfo {author} {\bibfnamefont {D.}~\bibnamefont {Ghazaryan}}, \bibinfo {author} {\bibfnamefont {A.}~\bibnamefont {Misra}}, \bibinfo {author} {\bibfnamefont {Y.}~\bibnamefont {Cao}}, \bibinfo {author} {\bibfnamefont {J.~S.}\ \bibnamefont {Tu}}, \bibinfo {author} {\bibfnamefont {B.~A.}\ \bibnamefont {Piot}}, \bibinfo {author} {\bibfnamefont {M.}~\bibnamefont {Potemski}}, \bibinfo {author} {\bibfnamefont {S.}~\bibnamefont {Pezzini}}, \bibinfo {author} {\bibfnamefont {S.}~\bibnamefont {Wiedmann}}, \bibinfo {author} {\bibfnamefont {U.}~\bibnamefont {Zeitler}}, \bibinfo {author} {\bibfnamefont {T.~L.~M.}\ \bibnamefont {Lane}}, \bibinfo {author} {\bibfnamefont {S.~V.}\ \bibnamefont {Morozov}}, \bibinfo {author} {\bibfnamefont {M.~T.}\ \bibnamefont {Greenaway}}, \bibinfo {author} {\bibfnamefont {L.}~\bibnamefont {Eaves}}, \bibinfo {author} {\bibfnamefont {A.~K.}\ \bibnamefont {Geim}}, \bibinfo {author} {\bibfnamefont
  {V.~I.}\ \bibnamefont {Fal'ko}}, \bibinfo {author} {\bibfnamefont {K.~S.}\ \bibnamefont {Novoselov}},\ and\ \bibinfo {author} {\bibfnamefont {A.}~\bibnamefont {Mishchenko}},\ }\bibfield  {title} {\bibinfo {title} {Tuning the valley and chiral quantum state of {D}irac electrons in van der {W}aals heterostructures},\ }\href {https://arxiv.org/pdf/1608.02411\#page=27} {\bibfield  {journal} {\bibinfo  {journal} {Science}\ }\textbf {\bibinfo {volume} {353}},\ \bibinfo {pages} {575} (\bibinfo {year} {2016})}\BibitemShut {NoStop}%
\bibitem [{\citenamefont {Feenstra}\ \emph {et~al.}(2012)\citenamefont {Feenstra}, \citenamefont {Jena},\ and\ \citenamefont {Gu}}]{tunneling_Gr_coherence_length}%
  \BibitemOpen
  \bibfield  {author} {\bibinfo {author} {\bibfnamefont {R.~M.}\ \bibnamefont {Feenstra}}, \bibinfo {author} {\bibfnamefont {D.}~\bibnamefont {Jena}},\ and\ \bibinfo {author} {\bibfnamefont {G.}~\bibnamefont {Gu}},\ }\bibfield  {title} {\bibinfo {title} {Single-particle tunneling in doped graphene-insulator-graphene junctions},\ }\href {https://arxiv.org/pdf/1108.4881\#page=8} {\bibfield  {journal} {\bibinfo  {journal} {Journal of Applied Physics}\ }\textbf {\bibinfo {volume} {111}},\ \bibinfo {pages} {043711} (\bibinfo {year} {2012})}\BibitemShut {NoStop}%
\bibitem [{\citenamefont {Katkov}\ and\ \citenamefont {Osipov}(2017)}]{Gr_tunneling_theory_review}%
  \BibitemOpen
  \bibfield  {author} {\bibinfo {author} {\bibfnamefont {V.~L.}\ \bibnamefont {Katkov}}\ and\ \bibinfo {author} {\bibfnamefont {V.~A.}\ \bibnamefont {Osipov}},\ }\bibfield  {title} {\bibinfo {title} {Review article: {T}unneling-based graphene electronics: {M}ethods and examples},\ }\href {https://pubs.aip.org/avs/jvb/article-pdf/doi/10.1116/1.4995380/9819186/050801\_1\_online.pdf\#page=12} {\bibfield  {journal} {\bibinfo  {journal} {Journal of Vacuum Science \& Technology B}\ }\textbf {\bibinfo {volume} {35}},\ \bibinfo {pages} {050801} (\bibinfo {year} {2017})}\BibitemShut {NoStop}%
\bibitem [{\citenamefont {Zheng}\ and\ \citenamefont {MacDonald}(1993)}]{tunneling_disorder_sqrt_divergence}%
  \BibitemOpen
  \bibfield  {author} {\bibinfo {author} {\bibfnamefont {L.}~\bibnamefont {Zheng}}\ and\ \bibinfo {author} {\bibfnamefont {A.~H.}\ \bibnamefont {MacDonald}},\ }\bibfield  {title} {\bibinfo {title} {Tunneling conductance between parallel two-dimensional electron systems},\ }\href {https://arxiv.org/pdf/cond-mat/9212016\#page=6} {\bibfield  {journal} {\bibinfo  {journal} {Phys. Rev. B}\ }\textbf {\bibinfo {volume} {47}},\ \bibinfo {pages} {10619} (\bibinfo {year} {1993})}\BibitemShut {NoStop}%
\bibitem [{\citenamefont {Amorim}\ \emph {et~al.}(2016)\citenamefont {Amorim}, \citenamefont {Ribeiro},\ and\ \citenamefont {Peres}}]{Gr-hBN_tunneling_model_3}%
  \BibitemOpen
  \bibfield  {author} {\bibinfo {author} {\bibfnamefont {B.}~\bibnamefont {Amorim}}, \bibinfo {author} {\bibfnamefont {R.~M.}\ \bibnamefont {Ribeiro}},\ and\ \bibinfo {author} {\bibfnamefont {N.~M.~R.}\ \bibnamefont {Peres}},\ }\bibfield  {title} {\bibinfo {title} {Multiple negative differential conductance regions and inelastic phonon assisted tunneling in graphene/{$h-$BN}/graphene structures},\ }\href {https://arxiv.org/pdf/1603.04446} {\bibfield  {journal} {\bibinfo  {journal} {Phys. Rev. B}\ }\textbf {\bibinfo {volume} {93}},\ \bibinfo {pages} {235403} (\bibinfo {year} {2016})}\BibitemShut {NoStop}%
\bibitem [{\citenamefont {Datta}(1995)}]{Datta}%
  \BibitemOpen
  \bibfield  {author} {\bibinfo {author} {\bibfnamefont {S.}~\bibnamefont {Datta}},\ }\href {https://books.google.com/books?id=28BC-ofEhvUC\&pg=321} {\emph {\bibinfo {title} {Electronic Transport in Mesoscopic Systems}}},\ Cambridge Studies in Semiconductor Physics and Microelectronic Engineering\ (\bibinfo  {publisher} {Cambridge University Press},\ \bibinfo {year} {1995})\BibitemShut {NoStop}%
\bibitem [{\citenamefont {Brey}(2014)}]{Gr-hBN_tunneling_model}%
  \BibitemOpen
  \bibfield  {author} {\bibinfo {author} {\bibfnamefont {L.}~\bibnamefont {Brey}},\ }\bibfield  {title} {\bibinfo {title} {Coherent tunneling and negative differential conductivity in a graphene/{$h$-BN}/graphene heterostructure},\ }\href {https://arxiv.org/pdf/1403.6073} {\bibfield  {journal} {\bibinfo  {journal} {Phys. Rev. Appl.}\ }\textbf {\bibinfo {volume} {2}},\ \bibinfo {pages} {014003} (\bibinfo {year} {2014})}\BibitemShut {NoStop}%
\bibitem [{\citenamefont {Guerrero-Becerra}\ \emph {et~al.}(2016)\citenamefont {Guerrero-Becerra}, \citenamefont {Tomadin},\ and\ \citenamefont {Polini}}]{Gr-hBN_tunneling_model_e-e}%
  \BibitemOpen
  \bibfield  {author} {\bibinfo {author} {\bibfnamefont {K.~A.}\ \bibnamefont {Guerrero-Becerra}}, \bibinfo {author} {\bibfnamefont {A.}~\bibnamefont {Tomadin}},\ and\ \bibinfo {author} {\bibfnamefont {M.}~\bibnamefont {Polini}},\ }\bibfield  {title} {\bibinfo {title} {Resonant tunneling and the quasiparticle lifetime in graphene/boron nitride/graphene heterostructures},\ }\href {https://arxiv.org/pdf/1512.08684\#page=6} {\bibfield  {journal} {\bibinfo  {journal} {Phys. Rev. B}\ }\textbf {\bibinfo {volume} {93}},\ \bibinfo {pages} {125417} (\bibinfo {year} {2016})}\BibitemShut {NoStop}%
\bibitem [{\citenamefont {Ge}\ \emph {et~al.}(2017)\citenamefont {Ge}, \citenamefont {Habib}, \citenamefont {De}, \citenamefont {Barlas}, \citenamefont {Wickramaratne}, \citenamefont {Neupane},\ and\ \citenamefont {Lake}}]{Gr-hBN_tunneling_model_4}%
  \BibitemOpen
  \bibfield  {author} {\bibinfo {author} {\bibfnamefont {S.}~\bibnamefont {Ge}}, \bibinfo {author} {\bibfnamefont {K.~M.~M.}\ \bibnamefont {Habib}}, \bibinfo {author} {\bibfnamefont {A.}~\bibnamefont {De}}, \bibinfo {author} {\bibfnamefont {Y.}~\bibnamefont {Barlas}}, \bibinfo {author} {\bibfnamefont {D.}~\bibnamefont {Wickramaratne}}, \bibinfo {author} {\bibfnamefont {M.~R.}\ \bibnamefont {Neupane}},\ and\ \bibinfo {author} {\bibfnamefont {R.~K.}\ \bibnamefont {Lake}},\ }\bibfield  {title} {\bibinfo {title} {Interlayer transport through a graphene/rotated boron nitride/graphene heterostructure},\ }\href {https://arxiv.org/pdf/1609.01369} {\bibfield  {journal} {\bibinfo  {journal} {Phys. Rev. B}\ }\textbf {\bibinfo {volume} {95}},\ \bibinfo {pages} {045303} (\bibinfo {year} {2017})}\BibitemShut {NoStop}%
\bibitem [{\citenamefont {Gurvitz}(1998)}]{multidot_Gurvitz}%
  \BibitemOpen
  \bibfield  {author} {\bibinfo {author} {\bibfnamefont {S.~A.}\ \bibnamefont {Gurvitz}},\ }\bibfield  {title} {\bibinfo {title} {Rate equations for quantum transport in multidot systems},\ }\href {https://arxiv.org/pdf/cond-mat/9702071\#page=7} {\bibfield  {journal} {\bibinfo  {journal} {Phys. Rev. B}\ }\textbf {\bibinfo {volume} {57}},\ \bibinfo {pages} {6602} (\bibinfo {year} {1998})}\BibitemShut {NoStop}%
\bibitem [{\citenamefont {Wegewijs}\ and\ \citenamefont {Nazarov}(1999)}]{multidot_Nazarov}%
  \BibitemOpen
  \bibfield  {author} {\bibinfo {author} {\bibfnamefont {M.~R.}\ \bibnamefont {Wegewijs}}\ and\ \bibinfo {author} {\bibfnamefont {Y.~V.}\ \bibnamefont {Nazarov}},\ }\bibfield  {title} {\bibinfo {title} {Resonant tunneling through linear arrays of quantum dots},\ }\href {https://arxiv.org/pdf/cond-mat/9806192\#page=9} {\bibfield  {journal} {\bibinfo  {journal} {Phys. Rev. B}\ }\textbf {\bibinfo {volume} {60}},\ \bibinfo {pages} {14318} (\bibinfo {year} {1999})}\BibitemShut {NoStop}%
\bibitem [{\citenamefont {Sprekeler}\ \emph {et~al.}(2004)\citenamefont {Sprekeler}, \citenamefont {Kie\ss{}lich}, \citenamefont {Wacker},\ and\ \citenamefont {Sch\"oll}}]{double_dot_Schoell}%
  \BibitemOpen
  \bibfield  {author} {\bibinfo {author} {\bibfnamefont {H.}~\bibnamefont {Sprekeler}}, \bibinfo {author} {\bibfnamefont {G.}~\bibnamefont {Kie\ss{}lich}}, \bibinfo {author} {\bibfnamefont {A.}~\bibnamefont {Wacker}},\ and\ \bibinfo {author} {\bibfnamefont {E.}~\bibnamefont {Sch\"oll}},\ }\bibfield  {title} {\bibinfo {title} {Coulomb effects in tunneling through a quantum dot stack},\ }\href {https://arxiv.org/pdf/cond-mat/0309696\#page=3} {\bibfield  {journal} {\bibinfo  {journal} {Phys. Rev. B}\ }\textbf {\bibinfo {volume} {69}},\ \bibinfo {pages} {125328} (\bibinfo {year} {2004})}\BibitemShut {NoStop}%
\bibitem [{\citenamefont {Oroszl\'any}\ \emph {et~al.}(2007)\citenamefont {Oroszl\'any}, \citenamefont {Korm\'anyos}, \citenamefont {Koltai}, \citenamefont {Cserti},\ and\ \citenamefont {Lambert}}]{double_dot_finite_T}%
  \BibitemOpen
  \bibfield  {author} {\bibinfo {author} {\bibfnamefont {L.}~\bibnamefont {Oroszl\'any}}, \bibinfo {author} {\bibfnamefont {A.}~\bibnamefont {Korm\'anyos}}, \bibinfo {author} {\bibfnamefont {J.}~\bibnamefont {Koltai}}, \bibinfo {author} {\bibfnamefont {J.}~\bibnamefont {Cserti}},\ and\ \bibinfo {author} {\bibfnamefont {C.~J.}\ \bibnamefont {Lambert}},\ }\bibfield  {title} {\bibinfo {title} {Nonthermal broadening in the conductance of double quantum dot structures},\ }\href {https://arxiv.org/pdf/cond-mat/0701656\#page=6} {\bibfield  {journal} {\bibinfo  {journal} {Phys. Rev. B}\ }\textbf {\bibinfo {volume} {76}},\ \bibinfo {pages} {045318} (\bibinfo {year} {2007})}\BibitemShut {NoStop}%
\bibitem [{\citenamefont {Yadalam}\ and\ \citenamefont {Harbola}(2018)}]{nonmonotonic_J_vs_gamma_3}%
  \BibitemOpen
  \bibfield  {author} {\bibinfo {author} {\bibfnamefont {H.~K.}\ \bibnamefont {Yadalam}}\ and\ \bibinfo {author} {\bibfnamefont {U.}~\bibnamefont {Harbola}},\ }\bibfield  {title} {\bibinfo {title} {Current in nanojunctions: {E}ffects of reservoir coupling},\ }\href {https://arxiv.org/pdf/1702.07122} {\bibfield  {journal} {\bibinfo  {journal} {Physica E: Low-dimensional Systems and Nanostructures}\ }\textbf {\bibinfo {volume} {101}},\ \bibinfo {pages} {224} (\bibinfo {year} {2018})}\BibitemShut {NoStop}%
\bibitem [{\citenamefont {Kazarinov}\ and\ \citenamefont {Suris}(1972)}]{superlattice_Kazarinov}%
  \BibitemOpen
  \bibfield  {author} {\bibinfo {author} {\bibfnamefont {R.}~\bibnamefont {Kazarinov}}\ and\ \bibinfo {author} {\bibfnamefont {R.}~\bibnamefont {Suris}},\ }\bibfield  {title} {\bibinfo {title} {Electric and electromagnetic properties of semiconductors with a superlattice},\ }\href {https://www.researchgate.net/profile/Ra-Suris/publication/257947852\_Electric\_and\_electromagnetic\_properties\_of\_semiconductors\_with\_a\_superlattice/links/0c960526bda1be0749000000/Electric-and-electromagnetic-properties-of-semiconductors-with-a-superlattice.pdf\#page=7} {\bibfield  {journal} {\bibinfo  {journal} {Sov. Phys. Semicond}\ }\textbf {\bibinfo {volume} {6}},\ \bibinfo {pages} {120} (\bibinfo {year} {1972})}\BibitemShut {NoStop}%
\bibitem [{\citenamefont {Capasso}\ \emph {et~al.}(1986)\citenamefont {Capasso}, \citenamefont {Mohammed},\ and\ \citenamefont {Cho}}]{superlattice_Capasso}%
  \BibitemOpen
  \bibfield  {author} {\bibinfo {author} {\bibfnamefont {F.}~\bibnamefont {Capasso}}, \bibinfo {author} {\bibfnamefont {K.}~\bibnamefont {Mohammed}},\ and\ \bibinfo {author} {\bibfnamefont {A.~Y.}\ \bibnamefont {Cho}},\ }\bibfield  {title} {\bibinfo {title} {Sequential resonant tunneling through a multiquantum well superlattice},\ }\href {https://doi.org/10.1063/1.97007} {\bibfield  {journal} {\bibinfo  {journal} {Applied Physics Letters}\ }\textbf {\bibinfo {volume} {48}},\ \bibinfo {pages} {478} (\bibinfo {year} {1986})}\BibitemShut {NoStop}%
\bibitem [{\citenamefont {Frishman}\ and\ \citenamefont {Gurvitz}(1993)}]{multiwell_Gurwitz}%
  \BibitemOpen
  \bibfield  {author} {\bibinfo {author} {\bibfnamefont {A.~M.}\ \bibnamefont {Frishman}}\ and\ \bibinfo {author} {\bibfnamefont {S.~A.}\ \bibnamefont {Gurvitz}},\ }\bibfield  {title} {\bibinfo {title} {Induced negative conductance in multiple-well heterostructures},\ }\href {https://doi.org/10.1103/PhysRevB.47.16348} {\bibfield  {journal} {\bibinfo  {journal} {Phys. Rev. B}\ }\textbf {\bibinfo {volume} {47}},\ \bibinfo {pages} {16348} (\bibinfo {year} {1993})}\BibitemShut {NoStop}%
\bibitem [{\citenamefont {Stafford}\ and\ \citenamefont {Wingreen}(1996)}]{double_well_Wingreen}%
  \BibitemOpen
  \bibfield  {author} {\bibinfo {author} {\bibfnamefont {C.~A.}\ \bibnamefont {Stafford}}\ and\ \bibinfo {author} {\bibfnamefont {N.~S.}\ \bibnamefont {Wingreen}},\ }\bibfield  {title} {\bibinfo {title} {Resonant photon-assisted tunneling through a double quantum dot: An electron pump from spatial {R}abi oscillations},\ }\href {https://arxiv.org/pdf/cond-mat/9509120\#page=4} {\bibfield  {journal} {\bibinfo  {journal} {Phys. Rev. Lett.}\ }\textbf {\bibinfo {volume} {76}},\ \bibinfo {pages} {1916} (\bibinfo {year} {1996})}\BibitemShut {NoStop}%
\bibitem [{\citenamefont {Nagase}\ \emph {et~al.}(2001)\citenamefont {Nagase}, \citenamefont {Furuya}, \citenamefont {Machida},\ and\ \citenamefont {Kurahashi}}]{double_well_Nagase}%
  \BibitemOpen
  \bibfield  {author} {\bibinfo {author} {\bibfnamefont {M.}~\bibnamefont {Nagase}}, \bibinfo {author} {\bibfnamefont {K.}~\bibnamefont {Furuya}}, \bibinfo {author} {\bibfnamefont {N.}~\bibnamefont {Machida}},\ and\ \bibinfo {author} {\bibfnamefont {M.}~\bibnamefont {Kurahashi}},\ }\bibfield  {title} {\bibinfo {title} {Current peak characteristics of triple-barrier resonant-tunneling diodes with and without phase breaking},\ }\href {https://doi.org/10.1143/JJAP.40.6753} {\bibfield  {journal} {\bibinfo  {journal} {Japanese Journal of Applied Physics}\ }\textbf {\bibinfo {volume} {40}},\ \bibinfo {pages} {6753} (\bibinfo {year} {2001})}\BibitemShut {NoStop}%
\bibitem [{\citenamefont {Zohta}\ \emph {et~al.}(1989)\citenamefont {Zohta}, \citenamefont {Nozu},\ and\ \citenamefont {Obara}}]{double_well_exp_splitting}%
  \BibitemOpen
  \bibfield  {author} {\bibinfo {author} {\bibfnamefont {Y.}~\bibnamefont {Zohta}}, \bibinfo {author} {\bibfnamefont {T.}~\bibnamefont {Nozu}},\ and\ \bibinfo {author} {\bibfnamefont {M.}~\bibnamefont {Obara}},\ }\bibfield  {title} {\bibinfo {title} {Resonant tunneling spectroscopy of two coupled quantum wells},\ }\href {https://doi.org/10.1103/PhysRevB.39.1375} {\bibfield  {journal} {\bibinfo  {journal} {Phys. Rev. B}\ }\textbf {\bibinfo {volume} {39}},\ \bibinfo {pages} {1375} (\bibinfo {year} {1989})}\BibitemShut {NoStop}%
\bibitem [{\citenamefont {Gruss}\ \emph {et~al.}(2016)\citenamefont {Gruss}, \citenamefont {Velizhanin},\ and\ \citenamefont {Zwolak}}]{nonmonotonic_J_vs_gamma_1}%
  \BibitemOpen
  \bibfield  {author} {\bibinfo {author} {\bibfnamefont {D.}~\bibnamefont {Gruss}}, \bibinfo {author} {\bibfnamefont {K.~A.}\ \bibnamefont {Velizhanin}},\ and\ \bibinfo {author} {\bibfnamefont {M.}~\bibnamefont {Zwolak}},\ }\bibfield  {title} {\bibinfo {title} {{L}andauer's formula with finite-time relaxation: {K}ramers' crossover in electronic transport},\ }\href {https://arxiv.org/pdf/1604.02962.pdf\#page=4} {\bibfield  {journal} {\bibinfo  {journal} {Scientific Reports}\ }\textbf {\bibinfo {volume} {6}},\ \bibinfo {pages} {24514} (\bibinfo {year} {2016})}\BibitemShut {NoStop}%
\bibitem [{\citenamefont {Gruss}\ \emph {et~al.}(2017)\citenamefont {Gruss}, \citenamefont {Smolyanitsky},\ and\ \citenamefont {Zwolak}}]{nonmonotonic_J_vs_gamma_2}%
  \BibitemOpen
  \bibfield  {author} {\bibinfo {author} {\bibfnamefont {D.}~\bibnamefont {Gruss}}, \bibinfo {author} {\bibfnamefont {A.}~\bibnamefont {Smolyanitsky}},\ and\ \bibinfo {author} {\bibfnamefont {M.}~\bibnamefont {Zwolak}},\ }\bibfield  {title} {\bibinfo {title} {{Communication: {R}elaxation-limited electronic currents in extended reservoir simulations}},\ }\href {https://arxiv.org/pdf/1707.06650.pdf\#page=3} {\bibfield  {journal} {\bibinfo  {journal} {The Journal of Chemical Physics}\ }\textbf {\bibinfo {volume} {147}},\ \bibinfo {pages} {141102} (\bibinfo {year} {2017})}\BibitemShut {NoStop}%
\bibitem [{\citenamefont {Caroli}\ \emph {et~al.}(1971)\citenamefont {Caroli}, \citenamefont {Combescot}, \citenamefont {Nozieres},\ and\ \citenamefont {Saint-James}}]{Caroli}%
  \BibitemOpen
  \bibfield  {author} {\bibinfo {author} {\bibfnamefont {C.}~\bibnamefont {Caroli}}, \bibinfo {author} {\bibfnamefont {R.}~\bibnamefont {Combescot}}, \bibinfo {author} {\bibfnamefont {P.}~\bibnamefont {Nozieres}},\ and\ \bibinfo {author} {\bibfnamefont {D.}~\bibnamefont {Saint-James}},\ }\bibfield  {title} {\bibinfo {title} {Direct calculation of the tunneling current},\ }\href {https://doi.org/10.1088/0022-3719/4/8/018} {\bibfield  {journal} {\bibinfo  {journal} {Journal of Physics C: Solid State Physics}\ }\textbf {\bibinfo {volume} {4}},\ \bibinfo {pages} {916} (\bibinfo {year} {1971})}\BibitemShut {NoStop}%
\bibitem [{\citenamefont {Britnell}\ \emph {et~al.}(2012{\natexlab{a}})\citenamefont {Britnell}, \citenamefont {Gorbachev}, \citenamefont {Jalil}, \citenamefont {Belle}, \citenamefont {Schedin}, \citenamefont {Mishchenko}, \citenamefont {Georgiou}, \citenamefont {Katsnelson}, \citenamefont {Eaves}, \citenamefont {Morozov}, \citenamefont {Peres}, \citenamefont {Leist}, \citenamefont {Geim}, \citenamefont {Novoselov},\ and\ \citenamefont {Ponomarenko}}]{power-law_T_vs_E}%
  \BibitemOpen
  \bibfield  {author} {\bibinfo {author} {\bibfnamefont {L.}~\bibnamefont {Britnell}}, \bibinfo {author} {\bibfnamefont {R.~V.}\ \bibnamefont {Gorbachev}}, \bibinfo {author} {\bibfnamefont {R.}~\bibnamefont {Jalil}}, \bibinfo {author} {\bibfnamefont {B.~D.}\ \bibnamefont {Belle}}, \bibinfo {author} {\bibfnamefont {F.}~\bibnamefont {Schedin}}, \bibinfo {author} {\bibfnamefont {A.}~\bibnamefont {Mishchenko}}, \bibinfo {author} {\bibfnamefont {T.}~\bibnamefont {Georgiou}}, \bibinfo {author} {\bibfnamefont {M.~I.}\ \bibnamefont {Katsnelson}}, \bibinfo {author} {\bibfnamefont {L.}~\bibnamefont {Eaves}}, \bibinfo {author} {\bibfnamefont {S.~V.}\ \bibnamefont {Morozov}}, \bibinfo {author} {\bibfnamefont {N.~M.~R.}\ \bibnamefont {Peres}}, \bibinfo {author} {\bibfnamefont {J.}~\bibnamefont {Leist}}, \bibinfo {author} {\bibfnamefont {A.~K.}\ \bibnamefont {Geim}}, \bibinfo {author} {\bibfnamefont {K.~S.}\ \bibnamefont {Novoselov}},\ and\ \bibinfo {author} {\bibfnamefont {L.~A.}\ \bibnamefont {Ponomarenko}},\ }\bibfield
  {title} {\bibinfo {title} {Field-effect tunneling transistor based on vertical graphene heterostructures},\ }\href {https://arxiv.org/pdf/1112.4999\#page=9} {\bibfield  {journal} {\bibinfo  {journal} {Science}\ }\textbf {\bibinfo {volume} {335}},\ \bibinfo {pages} {947} (\bibinfo {year} {2012}{\natexlab{a}})}\BibitemShut {NoStop}%
\bibitem [{\citenamefont {Bistritzer}\ and\ \citenamefont {MacDonald}(2010)}]{tunneling_Gr_correlated_disorder_divergence}%
  \BibitemOpen
  \bibfield  {author} {\bibinfo {author} {\bibfnamefont {R.}~\bibnamefont {Bistritzer}}\ and\ \bibinfo {author} {\bibfnamefont {A.~H.}\ \bibnamefont {MacDonald}},\ }\bibfield  {title} {\bibinfo {title} {Transport between twisted graphene layers},\ }\href {https://arxiv.org/pdf/1002.2983\#page=7} {\bibfield  {journal} {\bibinfo  {journal} {Phys. Rev. B}\ }\textbf {\bibinfo {volume} {81}},\ \bibinfo {pages} {245412} (\bibinfo {year} {2010})}\BibitemShut {NoStop}%
\bibitem [{\citenamefont {Kramers}(1940)}]{lonely_chemical_reference}%
  \BibitemOpen
  \bibfield  {author} {\bibinfo {author} {\bibfnamefont {H.}~\bibnamefont {Kramers}},\ }\bibfield  {title} {\bibinfo {title} {Brownian motion in a field of force and the diffusion model of chemical reactions},\ }\href {http://gu-statphys.org/media/mydocs/nesp_kramers.pdf\#page=16} {\bibfield  {journal} {\bibinfo  {journal} {Physica}\ }\textbf {\bibinfo {volume} {7}},\ \bibinfo {pages} {284} (\bibinfo {year} {1940})}\BibitemShut {NoStop}%
\bibitem [{\citenamefont {Bena}\ and\ \citenamefont {Montambaux}(2009)}]{graphene_Hamiltonian}%
  \BibitemOpen
  \bibfield  {author} {\bibinfo {author} {\bibfnamefont {C.}~\bibnamefont {Bena}}\ and\ \bibinfo {author} {\bibfnamefont {G.}~\bibnamefont {Montambaux}},\ }\bibfield  {title} {\bibinfo {title} {Remarks on the tight-binding model of graphene},\ }\href {https://arxiv.org/pdf/0712.0765\#page=7} {\bibfield  {journal} {\bibinfo  {journal} {New Journal of Physics}\ }\textbf {\bibinfo {volume} {11}},\ \bibinfo {pages} {095003} (\bibinfo {year} {2009})}\BibitemShut {NoStop}%
\bibitem [{\citenamefont {Ribeiro}\ and\ \citenamefont {Peres}(2011)}]{hBN_tight-binding}%
  \BibitemOpen
  \bibfield  {author} {\bibinfo {author} {\bibfnamefont {R.~M.}\ \bibnamefont {Ribeiro}}\ and\ \bibinfo {author} {\bibfnamefont {N.~M.~R.}\ \bibnamefont {Peres}},\ }\bibfield  {title} {\bibinfo {title} {Stability of boron nitride bilayers: {G}round-state energies, interlayer distances, and tight-binding description},\ }\href {https://arxiv.org/pdf/1101.3950\#page=3} {\bibfield  {journal} {\bibinfo  {journal} {Phys. Rev. B}\ }\textbf {\bibinfo {volume} {83}},\ \bibinfo {pages} {235312} (\bibinfo {year} {2011})}\BibitemShut {NoStop}%
\bibitem [{\citenamefont {Jung}\ \emph {et~al.}(2014)\citenamefont {Jung}, \citenamefont {Raoux}, \citenamefont {Qiao},\ and\ \citenamefont {MacDonald}}]{Gr-hBN_coupling}%
  \BibitemOpen
  \bibfield  {author} {\bibinfo {author} {\bibfnamefont {J.}~\bibnamefont {Jung}}, \bibinfo {author} {\bibfnamefont {A.}~\bibnamefont {Raoux}}, \bibinfo {author} {\bibfnamefont {Z.}~\bibnamefont {Qiao}},\ and\ \bibinfo {author} {\bibfnamefont {A.~H.}\ \bibnamefont {MacDonald}},\ }\bibfield  {title} {\bibinfo {title} {Ab initio theory of moir\'e superlattice bands in layered two-dimensional materials},\ }\href {https://arxiv.org/pdf/1312.7723\#page=10} {\bibfield  {journal} {\bibinfo  {journal} {Phys. Rev. B}\ }\textbf {\bibinfo {volume} {89}},\ \bibinfo {pages} {205414} (\bibinfo {year} {2014})}\BibitemShut {NoStop}%
\bibitem [{\citenamefont {Wallbank}\ \emph {et~al.}(2015)\citenamefont {Wallbank}, \citenamefont {Mucha-Kruczyński}, \citenamefont {Chen},\ and\ \citenamefont {Fal'ko}}]{Gr-hBN_coupling_2}%
  \BibitemOpen
  \bibfield  {author} {\bibinfo {author} {\bibfnamefont {J.~R.}\ \bibnamefont {Wallbank}}, \bibinfo {author} {\bibfnamefont {M.}~\bibnamefont {Mucha-Kruczyński}}, \bibinfo {author} {\bibfnamefont {X.}~\bibnamefont {Chen}},\ and\ \bibinfo {author} {\bibfnamefont {V.~I.}\ \bibnamefont {Fal'ko}},\ }\bibfield  {title} {\bibinfo {title} {Moiré superlattice effects in graphene/boron-nitride van der {W}aals heterostructures},\ }\href {https://arxiv.org/pdf/1411.1235\#page=9} {\bibfield  {journal} {\bibinfo  {journal} {Annalen der Physik}\ }\textbf {\bibinfo {volume} {527}},\ \bibinfo {pages} {359} (\bibinfo {year} {2015})}\BibitemShut {NoStop}%
\bibitem [{\citenamefont {Koshino}(2015)}]{twisted_coupling}%
  \BibitemOpen
  \bibfield  {author} {\bibinfo {author} {\bibfnamefont {M.}~\bibnamefont {Koshino}},\ }\bibfield  {title} {\bibinfo {title} {Interlayer interaction in general incommensurate atomic layers},\ }\href {https://iopscience.iop.org/article/10.1088/1367-2630/17/1/015014/pdf\#page=9} {\bibfield  {journal} {\bibinfo  {journal} {New Journal of Physics}\ }\textbf {\bibinfo {volume} {17}},\ \bibinfo {pages} {015014} (\bibinfo {year} {2015})}\BibitemShut {NoStop}%
\bibitem [{\citenamefont {Jones}(2003)}]{Gr_lattice_constant}%
  \BibitemOpen
  \bibfield  {author} {\bibinfo {author} {\bibfnamefont {L.}~\bibnamefont {Jones}},\ }\bibfield  {title} {\bibinfo {title} {Improved crystallographic data for graphite},\ }\href {https://www.researchgate.net/profile/Jane-Howe/publication/253038849\_Improved\_crystallographic\_data\_for\_graphite/links/54ebc3e60cf2082851be8f83/Improved-crystallographic-data-for-graphite.pdf} {\bibfield  {journal} {\bibinfo  {journal} {Powder diffraction}\ }\textbf {\bibinfo {volume} {18}},\ \bibinfo {pages} {150} (\bibinfo {year} {2003})}\BibitemShut {NoStop}%
\bibitem [{\citenamefont {Paszkowicz}\ \emph {et~al.}(2002)\citenamefont {Paszkowicz}, \citenamefont {Pelka}, \citenamefont {Knapp}, \citenamefont {Szyszko},\ and\ \citenamefont {Podsiadlo}}]{hBN_lattice_constant}%
  \BibitemOpen
  \bibfield  {author} {\bibinfo {author} {\bibfnamefont {W.}~\bibnamefont {Paszkowicz}}, \bibinfo {author} {\bibfnamefont {J.}~\bibnamefont {Pelka}}, \bibinfo {author} {\bibfnamefont {M.}~\bibnamefont {Knapp}}, \bibinfo {author} {\bibfnamefont {T.}~\bibnamefont {Szyszko}},\ and\ \bibinfo {author} {\bibfnamefont {S.}~\bibnamefont {Podsiadlo}},\ }\bibfield  {title} {\bibinfo {title} {Lattice parameters and anisotropic thermal expansion of hexagonal boron nitride in the 10--297.5 {K} temperature range},\ }\href {https://doi.org/10.1007/s003390100999} {\bibfield  {journal} {\bibinfo  {journal} {Applied Physics A}\ }\textbf {\bibinfo {volume} {75}},\ \bibinfo {pages} {431} (\bibinfo {year} {2002})}\BibitemShut {NoStop}%
\bibitem [{\citenamefont {Elias}\ \emph {et~al.}(2019)\citenamefont {Elias}, \citenamefont {Valvin}, \citenamefont {Pelini}, \citenamefont {Summerfield}, \citenamefont {Mellor}, \citenamefont {Cheng}, \citenamefont {Eaves}, \citenamefont {Foxon}, \citenamefont {Beton}, \citenamefont {Novikov}, \citenamefont {Gil},\ and\ \citenamefont {Cassabois}}]{mBN_direct_gap}%
  \BibitemOpen
  \bibfield  {author} {\bibinfo {author} {\bibfnamefont {C.}~\bibnamefont {Elias}}, \bibinfo {author} {\bibfnamefont {P.}~\bibnamefont {Valvin}}, \bibinfo {author} {\bibfnamefont {T.}~\bibnamefont {Pelini}}, \bibinfo {author} {\bibfnamefont {A.}~\bibnamefont {Summerfield}}, \bibinfo {author} {\bibfnamefont {C.~J.}\ \bibnamefont {Mellor}}, \bibinfo {author} {\bibfnamefont {T.~S.}\ \bibnamefont {Cheng}}, \bibinfo {author} {\bibfnamefont {L.}~\bibnamefont {Eaves}}, \bibinfo {author} {\bibfnamefont {C.~T.}\ \bibnamefont {Foxon}}, \bibinfo {author} {\bibfnamefont {P.~H.}\ \bibnamefont {Beton}}, \bibinfo {author} {\bibfnamefont {S.~V.}\ \bibnamefont {Novikov}}, \bibinfo {author} {\bibfnamefont {B.}~\bibnamefont {Gil}},\ and\ \bibinfo {author} {\bibfnamefont {G.}~\bibnamefont {Cassabois}},\ }\bibfield  {title} {\bibinfo {title} {Direct band-gap crossover in epitaxial monolayer boron nitride},\ }\href {https://hal.science/hal-02156532v1/file/s41467-019-10610-5.pdf\#page=4} {\bibfield  {journal} {\bibinfo  {journal}
  {Nature Communications}\ }\textbf {\bibinfo {volume} {10}},\ \bibinfo {pages} {2639} (\bibinfo {year} {2019})}\BibitemShut {NoStop}%
\bibitem [{\citenamefont {Cassabois}\ \emph {et~al.}(2016)\citenamefont {Cassabois}, \citenamefont {Valvin},\ and\ \citenamefont {Gil}}]{hBN_indirect_gap}%
  \BibitemOpen
  \bibfield  {author} {\bibinfo {author} {\bibfnamefont {G.}~\bibnamefont {Cassabois}}, \bibinfo {author} {\bibfnamefont {P.}~\bibnamefont {Valvin}},\ and\ \bibinfo {author} {\bibfnamefont {B.}~\bibnamefont {Gil}},\ }\bibfield  {title} {\bibinfo {title} {Hexagonal boron nitride is an indirect bandgap semiconductor},\ }\href {https://arxiv.org/pdf/1512.02962} {\bibfield  {journal} {\bibinfo  {journal} {Nature photonics}\ }\textbf {\bibinfo {volume} {10}},\ \bibinfo {pages} {262} (\bibinfo {year} {2016})}\BibitemShut {NoStop}%
\bibitem [{\citenamefont {Grenadier}\ \emph {et~al.}(2022)\citenamefont {Grenadier}, \citenamefont {Maity}, \citenamefont {Li}, \citenamefont {Lin},\ and\ \citenamefont {Jiang}}]{hBN_gap_photocurrent}%
  \BibitemOpen
  \bibfield  {author} {\bibinfo {author} {\bibfnamefont {S.~J.}\ \bibnamefont {Grenadier}}, \bibinfo {author} {\bibfnamefont {A.}~\bibnamefont {Maity}}, \bibinfo {author} {\bibfnamefont {J.}~\bibnamefont {Li}}, \bibinfo {author} {\bibfnamefont {J.}~\bibnamefont {Lin}},\ and\ \bibinfo {author} {\bibfnamefont {H.}~\bibnamefont {Jiang}},\ }\bibfield  {title} {\bibinfo {title} {Effects of unique band structure of {$h$-BN} probed by photocurrent excitation spectroscopy},\ }\href {https://iopscience.iop.org/article/10.35848/1882-0786/ac6b83/ampdf\#page=14} {\bibfield  {journal} {\bibinfo  {journal} {Applied Physics Express}\ }\textbf {\bibinfo {volume} {15}},\ \bibinfo {pages} {051005} (\bibinfo {year} {2022})}\BibitemShut {NoStop}%
\bibitem [{\citenamefont {Román}\ \emph {et~al.}(2021)\citenamefont {Román}, \citenamefont {Costa}, \citenamefont {Zobelli}, \citenamefont {Elias}, \citenamefont {Valvin}, \citenamefont {Cassabois}, \citenamefont {Gil}, \citenamefont {Summerfield}, \citenamefont {Cheng}, \citenamefont {Mellor}, \citenamefont {Beton}, \citenamefont {Novikov},\ and\ \citenamefont {Zagonel}}]{Gr-hBN_band_alignment}%
  \BibitemOpen
  \bibfield  {author} {\bibinfo {author} {\bibfnamefont {R.~J.~P.}\ \bibnamefont {Román}}, \bibinfo {author} {\bibfnamefont {F.~J. R.~C.}\ \bibnamefont {Costa}}, \bibinfo {author} {\bibfnamefont {A.}~\bibnamefont {Zobelli}}, \bibinfo {author} {\bibfnamefont {C.}~\bibnamefont {Elias}}, \bibinfo {author} {\bibfnamefont {P.}~\bibnamefont {Valvin}}, \bibinfo {author} {\bibfnamefont {G.}~\bibnamefont {Cassabois}}, \bibinfo {author} {\bibfnamefont {B.}~\bibnamefont {Gil}}, \bibinfo {author} {\bibfnamefont {A.}~\bibnamefont {Summerfield}}, \bibinfo {author} {\bibfnamefont {T.~S.}\ \bibnamefont {Cheng}}, \bibinfo {author} {\bibfnamefont {C.~J.}\ \bibnamefont {Mellor}}, \bibinfo {author} {\bibfnamefont {P.~H.}\ \bibnamefont {Beton}}, \bibinfo {author} {\bibfnamefont {S.~V.}\ \bibnamefont {Novikov}},\ and\ \bibinfo {author} {\bibfnamefont {L.~F.}\ \bibnamefont {Zagonel}},\ }\bibfield  {title} {\bibinfo {title} {Band gap measurements of monolayer {h-BN} and insights into carbon-related point defects},\ }\href
  {https://arxiv.org/pdf/2107.07950\#page=11} {\bibfield  {journal} {\bibinfo  {journal} {2D Materials}\ }\textbf {\bibinfo {volume} {8}},\ \bibinfo {pages} {044001} (\bibinfo {year} {2021})}\BibitemShut {NoStop}%
\bibitem [{\citenamefont {Hengsberger}\ \emph {et~al.}(2020)\citenamefont {Hengsberger}, \citenamefont {Leuenberger}, \citenamefont {Schuler}, \citenamefont {Roth},\ and\ \citenamefont {Muntwiler}}]{mBN_gap_ARPES}%
  \BibitemOpen
  \bibfield  {author} {\bibinfo {author} {\bibfnamefont {M.}~\bibnamefont {Hengsberger}}, \bibinfo {author} {\bibfnamefont {D.}~\bibnamefont {Leuenberger}}, \bibinfo {author} {\bibfnamefont {A.}~\bibnamefont {Schuler}}, \bibinfo {author} {\bibfnamefont {S.}~\bibnamefont {Roth}},\ and\ \bibinfo {author} {\bibfnamefont {M.}~\bibnamefont {Muntwiler}},\ }\bibfield  {title} {\bibinfo {title} {Dynamics of excited interlayer states in hexagonal boron nitride monolayers},\ }\href {https://www.zora.uzh.ch/id/eprint/186924/8/JPhysD\_hBN\_Review.pdf\#page=23} {\bibfield  {journal} {\bibinfo  {journal} {Journal of Physics D: Applied Physics}\ }\textbf {\bibinfo {volume} {53}},\ \bibinfo {pages} {203001} (\bibinfo {year} {2020})}\BibitemShut {NoStop}%
\bibitem [{\citenamefont {Carpenter}\ and\ \citenamefont {Kirby}(1982)}]{hBN_gap_R_vs_T}%
  \BibitemOpen
  \bibfield  {author} {\bibinfo {author} {\bibfnamefont {L.~G.}\ \bibnamefont {Carpenter}}\ and\ \bibinfo {author} {\bibfnamefont {P.~J.}\ \bibnamefont {Kirby}},\ }\bibfield  {title} {\bibinfo {title} {The electrical resistivity of boron nitride over the temperature range 700 degrees {C} to 1400 degrees {C}},\ }\href {https://doi.org/10.1088/0022-3727/15/7/009} {\bibfield  {journal} {\bibinfo  {journal} {Journal of Physics D: Applied Physics}\ }\textbf {\bibinfo {volume} {15}},\ \bibinfo {pages} {1143} (\bibinfo {year} {1982})}\BibitemShut {NoStop}%
\bibitem [{\citenamefont {Frederikse}\ \emph {et~al.}(1985)\citenamefont {Frederikse}, \citenamefont {Kahn}, \citenamefont {Dragoo},\ and\ \citenamefont {Hosler}}]{hBN_gap_R_vs_T_2}%
  \BibitemOpen
  \bibfield  {author} {\bibinfo {author} {\bibfnamefont {H.~P.~R.}\ \bibnamefont {Frederikse}}, \bibinfo {author} {\bibfnamefont {A.~H.}\ \bibnamefont {Kahn}}, \bibinfo {author} {\bibfnamefont {A.~L.}\ \bibnamefont {Dragoo}},\ and\ \bibinfo {author} {\bibfnamefont {W.~R.}\ \bibnamefont {Hosler}},\ }\bibfield  {title} {\bibinfo {title} {Electrical resistivity and microwave transmission of hexagonal boron nitride},\ }\href {https://doi.org/10.1111/j.1151-2916.1985.tb09650.x} {\bibfield  {journal} {\bibinfo  {journal} {Journal of the American Ceramic Society}\ }\textbf {\bibinfo {volume} {68}},\ \bibinfo {pages} {131} (\bibinfo {year} {1985})}\BibitemShut {NoStop}%
\bibitem [{\citenamefont {Yang}\ \emph {et~al.}(2019)\citenamefont {Yang}, \citenamefont {Liu}, \citenamefont {Zhou}, \citenamefont {Gao}, \citenamefont {Wang}, \citenamefont {Lv}, \citenamefont {Yuan}, \citenamefont {Jin}, \citenamefont {Zhao}, \citenamefont {Wei}, \citenamefont {Zhang}, \citenamefont {Gao}, \citenamefont {Li}, \citenamefont {Fan},\ and\ \citenamefont {Jiang}}]{hBN_gap_R_vs_T_3}%
  \BibitemOpen
  \bibfield  {author} {\bibinfo {author} {\bibfnamefont {X.}~\bibnamefont {Yang}}, \bibinfo {author} {\bibfnamefont {P.}~\bibnamefont {Liu}}, \bibinfo {author} {\bibfnamefont {D.}~\bibnamefont {Zhou}}, \bibinfo {author} {\bibfnamefont {F.}~\bibnamefont {Gao}}, \bibinfo {author} {\bibfnamefont {X.}~\bibnamefont {Wang}}, \bibinfo {author} {\bibfnamefont {S.}~\bibnamefont {Lv}}, \bibinfo {author} {\bibfnamefont {Z.}~\bibnamefont {Yuan}}, \bibinfo {author} {\bibfnamefont {X.}~\bibnamefont {Jin}}, \bibinfo {author} {\bibfnamefont {W.}~\bibnamefont {Zhao}}, \bibinfo {author} {\bibfnamefont {H.}~\bibnamefont {Wei}}, \bibinfo {author} {\bibfnamefont {L.}~\bibnamefont {Zhang}}, \bibinfo {author} {\bibfnamefont {J.}~\bibnamefont {Gao}}, \bibinfo {author} {\bibfnamefont {Q.}~\bibnamefont {Li}}, \bibinfo {author} {\bibfnamefont {S.}~\bibnamefont {Fan}},\ and\ \bibinfo {author} {\bibfnamefont {K.}~\bibnamefont {Jiang}},\ }\bibfield  {title} {\bibinfo {title} {High temperature performance of coaxial {h-BN}/{CNT} wires
  above 1,000 {\textdegree}{C}: {T}hermionic electron emission and thermally activated conductivity},\ }\href {https://doi.org/10.1007/s12274-019-2447-z} {\bibfield  {journal} {\bibinfo  {journal} {Nano Research}\ }\textbf {\bibinfo {volume} {12}},\ \bibinfo {pages} {1855} (\bibinfo {year} {2019})}\BibitemShut {NoStop}%
\bibitem [{\citenamefont {Pierucci}\ \emph {et~al.}(2018)\citenamefont {Pierucci}, \citenamefont {Zribi}, \citenamefont {Henck}, \citenamefont {Chaste}, \citenamefont {Silly}, \citenamefont {Bertran}, \citenamefont {Le~Fevre}, \citenamefont {Gil}, \citenamefont {Summerfield}, \citenamefont {Beton}, \citenamefont {Novikov}, \citenamefont {Cassabois}, \citenamefont {Rault},\ and\ \citenamefont {Ouerghi}}]{Gr-hBN_band_alignment_2}%
  \BibitemOpen
  \bibfield  {author} {\bibinfo {author} {\bibfnamefont {D.}~\bibnamefont {Pierucci}}, \bibinfo {author} {\bibfnamefont {J.}~\bibnamefont {Zribi}}, \bibinfo {author} {\bibfnamefont {H.}~\bibnamefont {Henck}}, \bibinfo {author} {\bibfnamefont {J.}~\bibnamefont {Chaste}}, \bibinfo {author} {\bibfnamefont {M.~G.}\ \bibnamefont {Silly}}, \bibinfo {author} {\bibfnamefont {F.}~\bibnamefont {Bertran}}, \bibinfo {author} {\bibfnamefont {P.}~\bibnamefont {Le~Fevre}}, \bibinfo {author} {\bibfnamefont {B.}~\bibnamefont {Gil}}, \bibinfo {author} {\bibfnamefont {A.}~\bibnamefont {Summerfield}}, \bibinfo {author} {\bibfnamefont {P.~H.}\ \bibnamefont {Beton}}, \bibinfo {author} {\bibfnamefont {S.~V.}\ \bibnamefont {Novikov}}, \bibinfo {author} {\bibfnamefont {G.}~\bibnamefont {Cassabois}}, \bibinfo {author} {\bibfnamefont {J.~E.}\ \bibnamefont {Rault}},\ and\ \bibinfo {author} {\bibfnamefont {A.}~\bibnamefont {Ouerghi}},\ }\bibfield  {title} {\bibinfo {title} {Van der {W}aals epitaxy of two-dimensional single-layer {h-BN}
  on graphite by molecular beam epitaxy: {E}lectronic properties and band structure},\ }\href {https://pubs.aip.org/aip/apl/article-pdf/doi/10.1063/1.5029220/14514004/253102\_1\_online.pdf\#page=4} {\bibfield  {journal} {\bibinfo  {journal} {Applied Physics Letters}\ }\textbf {\bibinfo {volume} {112}},\ \bibinfo {pages} {253102} (\bibinfo {year} {2018})}\BibitemShut {NoStop}%
\bibitem [{\citenamefont {Britnell}\ \emph {et~al.}(2012{\natexlab{b}})\citenamefont {Britnell}, \citenamefont {Gorbachev}, \citenamefont {Jalil}, \citenamefont {Belle}, \citenamefont {Schedin}, \citenamefont {Katsnelson}, \citenamefont {Eaves}, \citenamefont {Morozov}, \citenamefont {Mayorov}, \citenamefont {Peres}, \citenamefont {Castro~Neto}, \citenamefont {Leist}, \citenamefont {Geim}, \citenamefont {Ponomarenko},\ and\ \citenamefont {Novoselov}}]{current_vs_Nlayers}%
  \BibitemOpen
  \bibfield  {author} {\bibinfo {author} {\bibfnamefont {L.}~\bibnamefont {Britnell}}, \bibinfo {author} {\bibfnamefont {R.~V.}\ \bibnamefont {Gorbachev}}, \bibinfo {author} {\bibfnamefont {R.}~\bibnamefont {Jalil}}, \bibinfo {author} {\bibfnamefont {B.~D.}\ \bibnamefont {Belle}}, \bibinfo {author} {\bibfnamefont {F.}~\bibnamefont {Schedin}}, \bibinfo {author} {\bibfnamefont {M.~I.}\ \bibnamefont {Katsnelson}}, \bibinfo {author} {\bibfnamefont {L.}~\bibnamefont {Eaves}}, \bibinfo {author} {\bibfnamefont {S.~V.}\ \bibnamefont {Morozov}}, \bibinfo {author} {\bibfnamefont {A.~S.}\ \bibnamefont {Mayorov}}, \bibinfo {author} {\bibfnamefont {N.~M.~R.}\ \bibnamefont {Peres}}, \bibinfo {author} {\bibfnamefont {A.~H.}\ \bibnamefont {Castro~Neto}}, \bibinfo {author} {\bibfnamefont {J.}~\bibnamefont {Leist}}, \bibinfo {author} {\bibfnamefont {A.~K.}\ \bibnamefont {Geim}}, \bibinfo {author} {\bibfnamefont {L.~A.}\ \bibnamefont {Ponomarenko}},\ and\ \bibinfo {author} {\bibfnamefont {K.~S.}\ \bibnamefont {Novoselov}},\
  }\bibfield  {title} {\bibinfo {title} {Electron tunneling through ultrathin boron nitride crystalline barriers},\ }\href {https://repository.ubn.ru.nl/bitstream/handle/2066/103334/4/103334.pdf\#page=4} {\bibfield  {journal} {\bibinfo  {journal} {Nano Letters}\ }\textbf {\bibinfo {volume} {12}},\ \bibinfo {pages} {1707} (\bibinfo {year} {2012}{\natexlab{b}})}\BibitemShut {NoStop}%
\bibitem [{\citenamefont {Velický}\ \emph {et~al.}(2020)\citenamefont {Velický}, \citenamefont {Hu}, \citenamefont {Woods}, \citenamefont {Tóth}, \citenamefont {Zólyomi}, \citenamefont {Geim}, \citenamefont {Abruña}, \citenamefont {Novoselov},\ and\ \citenamefont {Dryfe}}]{current_vs_Nlayers_2}%
  \BibitemOpen
  \bibfield  {author} {\bibinfo {author} {\bibfnamefont {M.}~\bibnamefont {Velický}}, \bibinfo {author} {\bibfnamefont {S.}~\bibnamefont {Hu}}, \bibinfo {author} {\bibfnamefont {C.~R.}\ \bibnamefont {Woods}}, \bibinfo {author} {\bibfnamefont {P.~S.}\ \bibnamefont {Tóth}}, \bibinfo {author} {\bibfnamefont {V.}~\bibnamefont {Zólyomi}}, \bibinfo {author} {\bibfnamefont {A.~K.}\ \bibnamefont {Geim}}, \bibinfo {author} {\bibfnamefont {H.~D.}\ \bibnamefont {Abruña}}, \bibinfo {author} {\bibfnamefont {K.~S.}\ \bibnamefont {Novoselov}},\ and\ \bibinfo {author} {\bibfnamefont {R.~A.~W.}\ \bibnamefont {Dryfe}},\ }\bibfield  {title} {\bibinfo {title} {Electron tunneling through boron nitride confirms {M}arcus–{H}ush theory predictions for ultramicroelectrodes},\ }\href {https://pubs.acs.org/doi/epdf/10.1021/acsnano.9b08308\#page=5} {\bibfield  {journal} {\bibinfo  {journal} {ACS Nano}\ }\textbf {\bibinfo {volume} {14}},\ \bibinfo {pages} {993} (\bibinfo {year} {2020})}\BibitemShut {NoStop}%
\bibitem [{\citenamefont {Pratley}\ and\ \citenamefont {Z\"ulicke}(2013)}]{Gr_sqrt_divergence}%
  \BibitemOpen
  \bibfield  {author} {\bibinfo {author} {\bibfnamefont {L.}~\bibnamefont {Pratley}}\ and\ \bibinfo {author} {\bibfnamefont {U.}~\bibnamefont {Z\"ulicke}},\ }\bibfield  {title} {\bibinfo {title} {Magnetotunneling spectroscopy of chiral two-dimensional electron systems},\ }\href {https://arxiv.org/pdf/1308.6314\#page=5} {\bibfield  {journal} {\bibinfo  {journal} {Phys. Rev. B}\ }\textbf {\bibinfo {volume} {88}},\ \bibinfo {pages} {245412} (\bibinfo {year} {2013})}\BibitemShut {NoStop}%
\bibitem [{\citenamefont {Vasko}\ \emph {et~al.}(2000)\citenamefont {Vasko}, \citenamefont {Balev},\ and\ \citenamefont {Studart}}]{tunneling_interface_roughness}%
  \BibitemOpen
  \bibfield  {author} {\bibinfo {author} {\bibfnamefont {F.~T.}\ \bibnamefont {Vasko}}, \bibinfo {author} {\bibfnamefont {O.~G.}\ \bibnamefont {Balev}},\ and\ \bibinfo {author} {\bibfnamefont {N.}~\bibnamefont {Studart}},\ }\bibfield  {title} {\bibinfo {title} {Inhomogeneous broadening of tunneling conductance in double quantum wells},\ }\href {https://arxiv.org/pdf/cond-mat/0007401\#page=6} {\bibfield  {journal} {\bibinfo  {journal} {Phys. Rev. B}\ }\textbf {\bibinfo {volume} {62}},\ \bibinfo {pages} {12940} (\bibinfo {year} {2000})}\BibitemShut {NoStop}%
\bibitem [{\citenamefont {Jungwirth}\ and\ \citenamefont {MacDonald}(1996)}]{tunneling_e-e}%
  \BibitemOpen
  \bibfield  {author} {\bibinfo {author} {\bibfnamefont {T.}~\bibnamefont {Jungwirth}}\ and\ \bibinfo {author} {\bibfnamefont {A.~H.}\ \bibnamefont {MacDonald}},\ }\bibfield  {title} {\bibinfo {title} {Electron-electron interactions and two-dimensional--two-dimensional tunneling},\ }\href {https://arxiv.org/pdf/cond-mat/9603001\#page=6} {\bibfield  {journal} {\bibinfo  {journal} {Phys. Rev. B}\ }\textbf {\bibinfo {volume} {53}},\ \bibinfo {pages} {7403} (\bibinfo {year} {1996})}\BibitemShut {NoStop}%
\bibitem [{\citenamefont {Marinescu}\ \emph {et~al.}(2002)\citenamefont {Marinescu}, \citenamefont {Quinn},\ and\ \citenamefont {Giuliani}}]{tunneling_e-e_e-pl}%
  \BibitemOpen
  \bibfield  {author} {\bibinfo {author} {\bibfnamefont {D.~C.}\ \bibnamefont {Marinescu}}, \bibinfo {author} {\bibfnamefont {J.~J.}\ \bibnamefont {Quinn}},\ and\ \bibinfo {author} {\bibfnamefont {G.~F.}\ \bibnamefont {Giuliani}},\ }\bibfield  {title} {\bibinfo {title} {Tunneling between dissimilar quantum wells: {A} probe of the energy-dependent quasiparticle lifetime},\ }\href {https://scholar.archive.org/work/2vkzjwccfrghvplhnccglzja6e/access/wayback/https://tigerprints.clemson.edu/cgi/viewcontent.cgi?article=1181\&context=physastro\_pubs\#page=4} {\bibfield  {journal} {\bibinfo  {journal} {Phys. Rev. B}\ }\textbf {\bibinfo {volume} {65}},\ \bibinfo {pages} {045325} (\bibinfo {year} {2002})}\BibitemShut {NoStop}%
\bibitem [{\citenamefont {Banszerus}\ \emph {et~al.}(2016)\citenamefont {Banszerus}, \citenamefont {Schmitz}, \citenamefont {Engels}, \citenamefont {Goldsche}, \citenamefont {Watanabe}, \citenamefont {Taniguchi}, \citenamefont {Beschoten},\ and\ \citenamefont {Stampfer}}]{Gr_record_mean_free_path}%
  \BibitemOpen
  \bibfield  {author} {\bibinfo {author} {\bibfnamefont {L.}~\bibnamefont {Banszerus}}, \bibinfo {author} {\bibfnamefont {M.}~\bibnamefont {Schmitz}}, \bibinfo {author} {\bibfnamefont {S.}~\bibnamefont {Engels}}, \bibinfo {author} {\bibfnamefont {M.}~\bibnamefont {Goldsche}}, \bibinfo {author} {\bibfnamefont {K.}~\bibnamefont {Watanabe}}, \bibinfo {author} {\bibfnamefont {T.}~\bibnamefont {Taniguchi}}, \bibinfo {author} {\bibfnamefont {B.}~\bibnamefont {Beschoten}},\ and\ \bibinfo {author} {\bibfnamefont {C.}~\bibnamefont {Stampfer}},\ }\bibfield  {title} {\bibinfo {title} {Ballistic transport exceeding 28 $\mu$m in {CVD} grown graphene},\ }\href {https://arxiv.org/pdf/1511.08601\#page=7} {\bibfield  {journal} {\bibinfo  {journal} {Nano Letters}\ }\textbf {\bibinfo {volume} {16}},\ \bibinfo {pages} {1387} (\bibinfo {year} {2016})}\BibitemShut {NoStop}%
\bibitem [{\citenamefont {Hwang}\ and\ \citenamefont {Das~Sarma}(2008)}]{transport_vs_quasiparticle_relaxation_time}%
  \BibitemOpen
  \bibfield  {author} {\bibinfo {author} {\bibfnamefont {E.~H.}\ \bibnamefont {Hwang}}\ and\ \bibinfo {author} {\bibfnamefont {S.}~\bibnamefont {Das~Sarma}},\ }\bibfield  {title} {\bibinfo {title} {Single-particle relaxation time versus transport scattering time in a two-dimensional graphene layer},\ }\href {https://arxiv.org/pdf/0801.4736\#page=5} {\bibfield  {journal} {\bibinfo  {journal} {Phys. Rev. B}\ }\textbf {\bibinfo {volume} {77}},\ \bibinfo {pages} {195412} (\bibinfo {year} {2008})}\BibitemShut {NoStop}%
\bibitem [{\citenamefont {Tiras}\ \emph {et~al.}(2021)\citenamefont {Tiras}, \citenamefont {Ardali}, \citenamefont {Firat}, \citenamefont {Arslan},\ and\ \citenamefont {Ozbay}}]{Gr_quantum_lifetime_vs_substrate}%
  \BibitemOpen
  \bibfield  {author} {\bibinfo {author} {\bibfnamefont {E.}~\bibnamefont {Tiras}}, \bibinfo {author} {\bibfnamefont {S.}~\bibnamefont {Ardali}}, \bibinfo {author} {\bibfnamefont {H.}~\bibnamefont {Firat}}, \bibinfo {author} {\bibfnamefont {E.}~\bibnamefont {Arslan}},\ and\ \bibinfo {author} {\bibfnamefont {E.}~\bibnamefont {Ozbay}},\ }\bibfield  {title} {\bibinfo {title} {Substrate effects on electrical parameters of {D}irac fermions in graphene},\ }\href {https://repository.bilkent.edu.tr/bitstream/handle/11693/77558/Substrate\_effects\_on\_electrical\_parameters\_of\_Dirac\_fermions\_in\_graphene.pdf\#page=3} {\bibfield  {journal} {\bibinfo  {journal} {Materials Science in Semiconductor Processing}\ }\textbf {\bibinfo {volume} {133}},\ \bibinfo {pages} {105936} (\bibinfo {year} {2021})}\BibitemShut {NoStop}%
\bibitem [{\citenamefont {Li}\ and\ \citenamefont {Das~Sarma}(2013)}]{Gr_e-e_e-ph_rates}%
  \BibitemOpen
  \bibfield  {author} {\bibinfo {author} {\bibfnamefont {Q.}~\bibnamefont {Li}}\ and\ \bibinfo {author} {\bibfnamefont {S.}~\bibnamefont {Das~Sarma}},\ }\bibfield  {title} {\bibinfo {title} {Finite temperature inelastic mean free path and quasiparticle lifetime in graphene},\ }\href {https://arxiv.org/pdf/1211.6430\#page=3} {\bibfield  {journal} {\bibinfo  {journal} {Phys. Rev. B}\ }\textbf {\bibinfo {volume} {87}},\ \bibinfo {pages} {085406} (\bibinfo {year} {2013})}\BibitemShut {NoStop}%
\bibitem [{\citenamefont {Gilbertson}\ \emph {et~al.}(2012)\citenamefont {Gilbertson}, \citenamefont {Durakiewicz}, \citenamefont {Zhu}, \citenamefont {Mohite}, \citenamefont {Dattelbaum},\ and\ \citenamefont {Rodriguez}}]{Gr-hBN_SiO2_quantum_lifetime}%
  \BibitemOpen
  \bibfield  {author} {\bibinfo {author} {\bibfnamefont {S.}~\bibnamefont {Gilbertson}}, \bibinfo {author} {\bibfnamefont {T.}~\bibnamefont {Durakiewicz}}, \bibinfo {author} {\bibfnamefont {J.-X.}\ \bibnamefont {Zhu}}, \bibinfo {author} {\bibfnamefont {A.~D.}\ \bibnamefont {Mohite}}, \bibinfo {author} {\bibfnamefont {A.}~\bibnamefont {Dattelbaum}},\ and\ \bibinfo {author} {\bibfnamefont {G.}~\bibnamefont {Rodriguez}},\ }\bibfield  {title} {\bibinfo {title} {Direct measurement of quasiparticle lifetimes in graphene using time-resolved photoemission},\ }\href {https://www.researchgate.net/profile/Tomasz-Durakiewicz/publication/233756556\_Direct\_measurement\_of\_quasiparticle\_lifetimes\_in\_graphene\_using\_time-resolved\_photoemission/links/0fcfd512646eded5e4000000/Direct-measurement-of-quasiparticle-lifetimes-in-graphene-using-time-resolved-photoemission.pdf\#page=7} {\bibfield  {journal} {\bibinfo  {journal} {Journal of Vacuum Science \& Technology B}\ }\textbf {\bibinfo {volume} {30}},\ \bibinfo {pages}
  {03D116} (\bibinfo {year} {2012})}\BibitemShut {NoStop}%
\bibitem [{\citenamefont {Zeng}\ \emph {et~al.}(2019)\citenamefont {Zeng}, \citenamefont {Li}, \citenamefont {Dietrich}, \citenamefont {Ghosh}, \citenamefont {Watanabe}, \citenamefont {Taniguchi}, \citenamefont {Hone},\ and\ \citenamefont {Dean}}]{Gr-hBN_quantum_lifetime}%
  \BibitemOpen
  \bibfield  {author} {\bibinfo {author} {\bibfnamefont {Y.}~\bibnamefont {Zeng}}, \bibinfo {author} {\bibfnamefont {J.~I.~A.}\ \bibnamefont {Li}}, \bibinfo {author} {\bibfnamefont {S.~A.}\ \bibnamefont {Dietrich}}, \bibinfo {author} {\bibfnamefont {O.~M.}\ \bibnamefont {Ghosh}}, \bibinfo {author} {\bibfnamefont {K.}~\bibnamefont {Watanabe}}, \bibinfo {author} {\bibfnamefont {T.}~\bibnamefont {Taniguchi}}, \bibinfo {author} {\bibfnamefont {J.}~\bibnamefont {Hone}},\ and\ \bibinfo {author} {\bibfnamefont {C.~R.}\ \bibnamefont {Dean}},\ }\bibfield  {title} {\bibinfo {title} {High-quality magnetotransport in graphene using the edge-free {C}orbino geometry},\ }\href {https://arxiv.org/pdf/1805.04904.pdf\#page=2} {\bibfield  {journal} {\bibinfo  {journal} {Phys. Rev. Lett.}\ }\textbf {\bibinfo {volume} {122}},\ \bibinfo {pages} {137701} (\bibinfo {year} {2019})}\BibitemShut {NoStop}%
\bibitem [{\citenamefont {Spataru}\ and\ \citenamefont {Léonard}(2023)}]{BLG_Gr_quantum_lifetime_ab_initio}%
  \BibitemOpen
  \bibfield  {author} {\bibinfo {author} {\bibfnamefont {C.~D.}\ \bibnamefont {Spataru}}\ and\ \bibinfo {author} {\bibfnamefont {F.}~\bibnamefont {Léonard}},\ }\bibfield  {title} {\bibinfo {title} {Ab initio calculations of low-energy quasiparticle lifetimes in bilayer graphene},\ }\href {https://arxiv.org/pdf/2309.04048\#page=13} {\bibfield  {journal} {\bibinfo  {journal} {Applied Physics Letters}\ }\textbf {\bibinfo {volume} {123}},\ \bibinfo {pages} {113101} (\bibinfo {year} {2023})}\BibitemShut {NoStop}%
\bibitem [{\citenamefont {Borysenko}\ \emph {et~al.}(2010)\citenamefont {Borysenko}, \citenamefont {Mullen}, \citenamefont {Barry}, \citenamefont {Paul}, \citenamefont {Semenov}, \citenamefont {Zavada}, \citenamefont {Nardelli},\ and\ \citenamefont {Kim}}]{Gr_e-ph_rate}%
  \BibitemOpen
  \bibfield  {author} {\bibinfo {author} {\bibfnamefont {K.~M.}\ \bibnamefont {Borysenko}}, \bibinfo {author} {\bibfnamefont {J.~T.}\ \bibnamefont {Mullen}}, \bibinfo {author} {\bibfnamefont {E.~A.}\ \bibnamefont {Barry}}, \bibinfo {author} {\bibfnamefont {S.}~\bibnamefont {Paul}}, \bibinfo {author} {\bibfnamefont {Y.~G.}\ \bibnamefont {Semenov}}, \bibinfo {author} {\bibfnamefont {J.~M.}\ \bibnamefont {Zavada}}, \bibinfo {author} {\bibfnamefont {M.~B.}\ \bibnamefont {Nardelli}},\ and\ \bibinfo {author} {\bibfnamefont {K.~W.}\ \bibnamefont {Kim}},\ }\bibfield  {title} {\bibinfo {title} {First-principles analysis of electron-phonon interactions in graphene},\ }\href {https://arxiv.org/pdf/0912.0562\#page=12} {\bibfield  {journal} {\bibinfo  {journal} {Phys. Rev. B}\ }\textbf {\bibinfo {volume} {81}},\ \bibinfo {pages} {121412(R)} (\bibinfo {year} {2010})}\BibitemShut {NoStop}%
\bibitem [{\citenamefont {Woods}\ \emph {et~al.}(2014)\citenamefont {Woods}, \citenamefont {Britnell}, \citenamefont {Eckmann}, \citenamefont {Ma}, \citenamefont {Lu}, \citenamefont {Guo}, \citenamefont {Lin}, \citenamefont {Yu}, \citenamefont {Cao}, \citenamefont {Gorbachev}, \citenamefont {Kretinin}, \citenamefont {Park}, \citenamefont {Ponomarenko}, \citenamefont {Katsnelson}, \citenamefont {Gornostyrev}, \citenamefont {Watanabe}, \citenamefont {Taniguchi}, \citenamefont {Casiraghi}, \citenamefont {Gao}, \citenamefont {Geim},\ and\ \citenamefont {Novoselov}}]{Gr_commensurate-incommensurate}%
  \BibitemOpen
  \bibfield  {author} {\bibinfo {author} {\bibfnamefont {C.~R.}\ \bibnamefont {Woods}}, \bibinfo {author} {\bibfnamefont {L.}~\bibnamefont {Britnell}}, \bibinfo {author} {\bibfnamefont {A.}~\bibnamefont {Eckmann}}, \bibinfo {author} {\bibfnamefont {R.~S.}\ \bibnamefont {Ma}}, \bibinfo {author} {\bibfnamefont {J.~C.}\ \bibnamefont {Lu}}, \bibinfo {author} {\bibfnamefont {H.~M.}\ \bibnamefont {Guo}}, \bibinfo {author} {\bibfnamefont {X.}~\bibnamefont {Lin}}, \bibinfo {author} {\bibfnamefont {G.~L.}\ \bibnamefont {Yu}}, \bibinfo {author} {\bibfnamefont {Y.}~\bibnamefont {Cao}}, \bibinfo {author} {\bibfnamefont {R.}~\bibnamefont {Gorbachev}}, \bibinfo {author} {\bibfnamefont {A.~V.}\ \bibnamefont {Kretinin}}, \bibinfo {author} {\bibfnamefont {J.}~\bibnamefont {Park}}, \bibinfo {author} {\bibfnamefont {L.~A.}\ \bibnamefont {Ponomarenko}}, \bibinfo {author} {\bibfnamefont {M.~I.}\ \bibnamefont {Katsnelson}}, \bibinfo {author} {\bibfnamefont {Y.}~\bibnamefont {Gornostyrev}}, \bibinfo {author} {\bibfnamefont
  {K.}~\bibnamefont {Watanabe}}, \bibinfo {author} {\bibfnamefont {T.}~\bibnamefont {Taniguchi}}, \bibinfo {author} {\bibfnamefont {C.}~\bibnamefont {Casiraghi}}, \bibinfo {author} {\bibfnamefont {H.-J.}\ \bibnamefont {Gao}}, \bibinfo {author} {\bibfnamefont {A.~K.}\ \bibnamefont {Geim}},\ and\ \bibinfo {author} {\bibfnamefont {K.}~\bibnamefont {Novoselov}},\ }\bibfield  {title} {\bibinfo {title} {Commensurate--incommensurate transition in graphene on hexagonal boron nitride},\ }\href {https://arxiv.org/pdf/1401.2637\#page=3} {\bibfield  {journal} {\bibinfo  {journal} {Nature Physics}\ }\textbf {\bibinfo {volume} {10}},\ \bibinfo {pages} {451} (\bibinfo {year} {2014})}\BibitemShut {NoStop}%
\bibitem [{\citenamefont {{GNU TeXmacs}}()}]{TeXmacs}%
  \BibitemOpen
  \bibfield  {author} {\bibinfo {author} {\bibnamefont {{GNU TeXmacs}}},\ }\href {https://www.texmacs.org/tmweb/home/videos.en.html} {\bibinfo {title} {https://www.texmacs.org/}}\BibitemShut {NoStop}%
\end{thebibliography}
